\documentclass[aps,reprint,nofootinbib]{revtex4-1}

\usepackage[english]{babel}
\usepackage[utf8x]{inputenc}
\usepackage[T1]{fontenc}

\usepackage{amsmath}
\usepackage{hyperref}
\usepackage{cleveref}
\usepackage{graphicx}
\usepackage{glossaries}

\usepackage{physics}
\usepackage{siunitx}

\newcommand{\delay}[1]{\mathcal{D}_{#1}}
\newcommand{\filter}{\mathcal{F}}

\newcommand{\convolve}[2]{\qty(#1 \ast #2)}
\newcommand{\fourier}[1]{\opbraces{\widetilde{#1}}}

\newcommand{\psd}[1]{\opbraces{S_{#1}}}

\renewcommand{\qc}{\,\text{,}}
\newcommand{\qs}{\,\text{.}}

\newacronym{lisa}{LISA}{Laser Interferometer Space Antenna}
\newacronym{gw}{GW}{gravitational wave}
\newacronym{gpu}{GPU}{graphics processing unit}
\newacronym{esa}{ESA}{European Space Agency}
\newacronym{tdi}{TDI}{time-delay interferometry}
\newacronym{fir}{FIR}{finite impulse response}
\newacronym{iir}{IIR}{infinite impulse response}
\newacronym[longplural={power spectral densities}]{psd}{PSD}{power spectral density}
\newacronym[longplural={amplitude spectral densities}]{asd}{ASD}{amplitude spectral density}
\newacronym{ldc}{LDC}{LISA Data Challenge}
\newacronym{mldc}{MLDC}{Mock LISA Data Challenge}
\newacronym{ssb}{SSB}{Solar system's barycenter}
\newacronym{uflis}{UFLIS}{University of Florida LISA simulator}
\newacronym{lot}{LOT}{LISA On Table}
\newacronym{apc}{APC}{Astroparticules et Cosmologie}
\newacronym[longplural={movable optical system assemblies}]{mosa}{MOSA}{movable optical system assembly}

\begin{document}

\title{Effect of filters on the time-delay interferometry residual laser noise for LISA}

\author{Jean-Baptiste Bayle}
\author{Marc Lilley}
\author{Antoine Petiteau}
\author{Hubert Halloin}
\affiliation{APC, Université Paris Diderot, CNRS/IN2P3, CEA/Irfu, Observatoire de Paris, Sorbonne Paris Cité, 10 rue Alice Domont et Léonie Duquet, 75013 Paris, France}

\date{\today}


\begin{abstract}
The Laser Interferometer Space Antenna (LISA) is a European Space Agency mission that aims to measure gravitational waves in the millihertz range. Laser frequency noise enters the interferometric measurements and dominates the expected gravitational signals by many orders of magnitude. Time-delay interferometry (TDI) is a technique that reduces this laser noise by synthesizing virtual equal-arm interferometric measurements. Laboratory experiments and numerical simulations have confirmed that this reduction is sufficient to meet the scientific goals of the mission in proof-of-concept setups. In this paper, we show that the on-board antialiasing filters play an important role in TDI's performance when the flexing of the constellation is accounted for. This coupling was neglected in previous studies. To reach an optimal reduction level, filters with vanishing group delays must be used on board or synthesized off-line. We propose a theoretical model of the residual laser noise including this flexing-filtering coupling. We also use two independent simulators to produce realistic measurement signals and compute the corresponding TDI Michelson variables. We show that our theoretical model agrees with the simulated data with exquisite precision. Using these two complementary approaches, we confirm TDI's ability to reduce laser frequency noise in a more realistic mission setup. The theoretical model provides insight on filter design and implementation.
\end{abstract}
\maketitle

\section{Introduction}
\label{sec:introduction}

The \gls{lisa} is a \gls{esa} scientific space mission which aims to measure \glspl{gw} in the millihertz range~\cite{Audley:2017drz}. Those waves are predicted by Einstein's theory of general relativity and produced by the quadrupolar moment of very dense objects, such as black hole binaries or coalescing supermassive black holes. The detection of low-frequency gravitational waves will help answer numerous astrophysical, cosmological, and theoretical questions, related, for example, to the formation of black hole binaries and extreme mass ratio inspirals, the formation of galaxies or general relativity in the strong field regime~\cite{Audley:2017drz}.

The mission is expected to be launched in the year 2034. Three spacecraft will trail the Earth around the Sun, in a nearly equilateral triangular configuration with armlengths of about 2.5 million kilometers. Each spacecraft contains two free falling test masses acting as inertial sensors~\cite{Audley:2017drz} and two optical benches. Six laser links connect the six optical benches performing interferometric measurements between the local and distant laser beams. These optical setups are capable of measuring the differential displacement between the local and remote test masses with subpicometer precision~\cite{Armano:2016bkm,Armano:2018hg}. In the latest design, each spacecraft performs six interferometric measurements (see \cref{sec:instrumental-setup}), which are then telemetered to Earth.

Among the multiple sources of noise which enter the measurements made by \gls{lisa}, laser frequency noise is dominant. Its amplitude is greater than that of other (secondary) noises and that of GWs by several orders of magnitude. The armlengths of the \gls{lisa} constellation are indeed not equal, preventing laser noise to be canceled when the beams are recombined. \Gls{tdi} is an algorithm first proposed by~\cite{Armstrong:1999hp} that aims to reduce the laser frequency noise by 8 orders of magnitude, bringing it below secondary noises and \gls{gw} signals~\cite{Tinto:1999yr}. \Gls{tdi} synthesizes virtual equal-arm interferometric measurements by combining time-shifted measurements from \gls{lisa}.

Laboratory experiments have been performed to study whether \gls{tdi} can be applied correctly and whether its performance meets mission requirements, in various setups. A first demonstrator was designed at the Jet Propulsion Lab~\cite{deVine:2010et} to reproduce noise couplings and measure \gls{tdi} noise reduction in a fixed two-arm configuration. It was shown that the laser frequency noise could be reduced to the desired level. The Hexagon interferometer~\cite{Schwarze:2018lvl} is a metrology test bed developed at the Albert Einstein Institute, and consists of three locked lasers~\cite{Otto:2015wp}. It was used both to test the performance and feasibility of \gls{tdi} for heterodyne interferometry and to test phasemeter prototypes. \Gls{lot} is an electro-optical simulator developed at the laboratoire \gls{apc}~\cite{Laporte:2017bv,Gruning:2015cp}. The results show that in the case of two static unequal arms, first-generation \gls{tdi} cancels the laser frequency noise as expected. The \gls{uflis} uses electronic phase delay units to simulate the time variation of the armlengths. First-generation \gls{tdi} was successfully tested, reaching the required performance, while \gls{tdi} 2.0 results were limited by the noise of the electronic phase delay units~\cite{Cruz:2006js}. To date, however, no realistic demonstrator using time-varying armlengths was successful in measuring the performance of \gls{tdi} 2.0.

Computer simulations have also been used to check the performance of the laser frequency noise removal performed by \gls{tdi}. \texttt{Synthetic LISA} was developed by M. Vallisneri~\cite{Vallisneri:2005ca} to study \gls{tdi} laser noise reduction for a flexing constellation, in an idealized configuration. Using this simulator, it was shown that one must use the second-generation \gls{tdi} algorithms to meet noise reduction requirements. \texttt{LISACode}~\cite{Petiteau:2008wa,Petiteau:2008ke} is the tool currently used by the \gls{lisa} Simulation Group and the \gls{ldc} to produce realistic data. It uses a high-level model of the instrument to reproduce the instrumental response to incoming gravitational waves. \texttt{LISACode} also includes models for several sources of noise, including the aforementioned laser frequency noise. \texttt{LISANode}~\cite{Bayle:tbp} is a new prototype end-to-end mission simulator. It is a very flexible framework that enables the study of various instrumental configurations. \texttt{LISANode} includes an up-to-date model for the instrument, various sources of noise and the \gls{tdi} algorithms.

In this paper, we have developed an analytic model that describes both \gls{lisa} and \gls{tdi} for a realistic setup in order to determine which instrumental factors play an important role in the laser frequency noise reduction performance. This model reproduces the results of both \texttt{LISANode} and \texttt{LISACode} with great precision. We include the effect of the so-called flexing, \textit{i.e.}~time-varying armlengths, and that of the antialiasing filtering applied before the high-frequency measurements are downsampled and telemetered to Earth. The agreement between our theoretical model and our simulations demonstrates that \texttt{LISACode} and \texttt{LISANode} are implemented correctly. Our work also shows that a coupling between the flexing of the constellation and the antialiasing filters can degrade significantly \gls{tdi}'s performance. We show that the effect of this coupling is mitigated with well-designed filters and an specific off-line treatment. The remaining laser frequency noise can then be maintained below mission requirements in the frequency band between \num{E-4} and \SI{E-1}{Hz}, for second-generation \gls{tdi}.

We first present the \gls{lisa} mission setup that was modeled analytically and simulated numerically in \cref{sec:instrumental-setup}, and the \gls{tdi} algorithm that was used in \cref{sec:tdi}. In \cref{sec:analytic-model}, we derive the corresponding analytic model for \gls{tdi} 1.5 and 2.0. In \cref{sec:simulation} we describe \texttt{LISACode} and \texttt{LISANode}, and give details about the configuration used to generate the data. Finally, in \cref{sec:results}, we compare and discuss the results of the simulators and of the analytic approach. In particular, we study the effect of different types of filters and discuss their potential implementations.

\section{Instrumental setup}
\label{sec:instrumental-setup}

\begin{figure}
    \centering
    \includegraphics[width=\columnwidth]{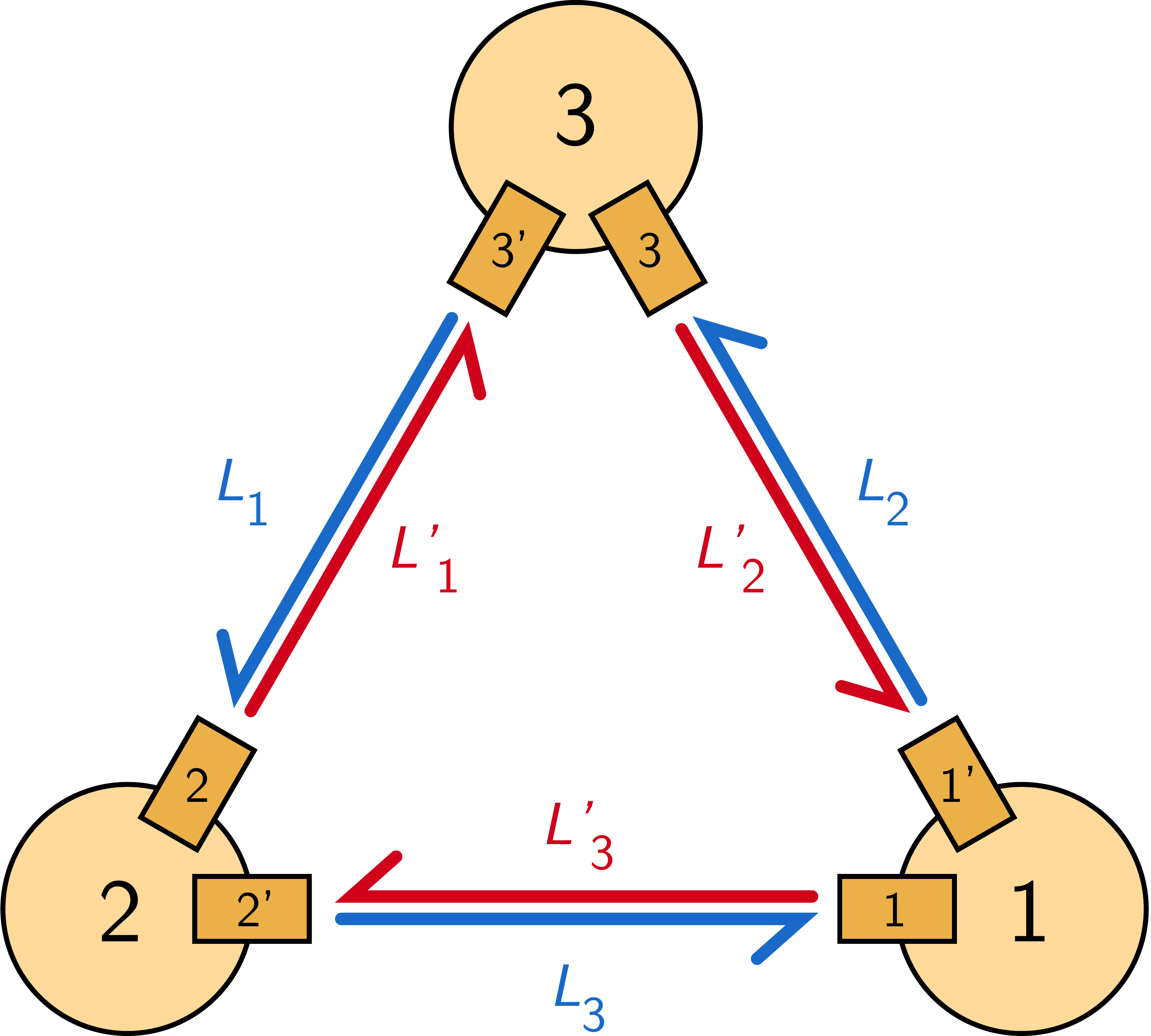}
    \caption{Conventions for labeling spacecraft, \glspl{mosa}, lasers, optical benches and arms. Primed indices are used for arms pointing clockwise, and for \glspl{mosa} and optical benches receiving light clockwise.}
    \label{fig:conventions}
\end{figure}

\Gls{lisa} is a constellation of three spacecraft forming a nearly equilateral triangle. The constellation's center of mass trails the Earth in its orbit around the Sun by around 20 deg.~\cite{Audley:2017drz}. Each spacecraft emits and receives a laser beam along each of the two arms connecting it to its companion spacecraft. All spacecraft host two \glspl{mosa}, an on-board computer, a phasemeter, and two laser sources. A \gls{mosa} is composed of a telescope and an optical bench. The telescope sends an outgoing laser beam to its companion spacecraft and collects incoming light. Various conventions are used in the literature to denote the spacecraft, optical benches and arms. In this paper, we number these components according to \cref{fig:conventions}.

In this paper, we consider the latest optical design, often called \textit{split interferometry}. It is extensively described in~\cite{Otto:2012dk}. Three interferometric measurements are performed on each optical bench $i$: the science $s_i$ (respectively, reference $\tau_i$) signal is the beat note between the distant (respectively, adjacent) and local beams without any reflection on the test masses. The test mass signal $\epsilon_i$ corresponds to the beat note formed by the local and adjacent beams after reflection onto the local test mass.

For simplicity, we neglect all secondary noise sources, such as the read-out noise, the optical path noise, and the test mass acceleration noise. The clocks on board each spacecraft are assumed to be perfect and we neglect both the clock noise and the sideband measurements that are used to remove parts of this noise~\cite{Tinto:2018kij,Otto:2012dk}. We therefore only study laser frequency noise, denoted $p_i(t)$ and $p_i'(t)$, where $i$ is the spacecraft index. We model it here as a white noise with a \gls{psd} of \SI{E-26}{Hz^{-1}}. Under those assumptions, \textit{split interferometry} reduces to the legacy design described in~\cite{Petiteau:2008ke}.

The interferometric measurements are delivered by the phasemeter to the on-board computer at \SI{20}{Hz}. Due to limitations in the telemetry passband, they must be downsampled to \SI{2}{Hz} before they are transmitted to Earth~\cite{Audley:2017drz}. Antialiasing filters are used to prevent power folding in the band of interest during decimation. These filters are assumed to be identical on board all spacecraft, and consist of a convolution with a filter kernel $f(t)$. We define the filter operator $\filter$, such that $\filter x(t) = \convolve{f}{x}(t)$ for any signal $x(t)$.

We model all signals as Doppler observables, \textit{i.e.}~as the ratio of the instantaneous frequency deviation from the nominal carrier frequency $\qty[\nu(t) - \nu_0] / \nu_0$~\cite{Vallisneri:2005ca} over that nominal carrier frequency. We neglect frequency planning~\cite{Barke:2015wb} and Doppler shifts due to the relative motion of the spacecraft. Therefore, in this paper, the carrier nominal frequency remains constant and equal for all beams, and interferometric signals are obtained by forming the difference of two incoming Doppler observables.

The propagation of laser beams between two spacecraft is modeled by applying time-varying delays. These delays correspond to the sum of all delays in the optical, analog and digital signal chains, though the main contribution remains the light travel times between the spacecraft. Therefore we suppose here that they are given by the armlengths and the speed of light in vacuum. We denote $\delay{i}$ the operator associated with traveling along arm $i$, of length $c \times L_i(t)$. For example, the laser frequency noise received by optical bench $2'$ from laser $1$, after it has traveled along arm $3$, is given by
\begin{equation}
    \delay{3} p_1 (t) = p_1 \qty(t - L_3(t)) \qs
    \label{eq:definition-delay-operator}
\end{equation}

We give the expressions of the measurement signals for the spacecraft 1; others are obtained by circular permutation of the indices. The science signals read
\begin{equation}
    \begin{aligned}
        s_1 &= \filter \delay{3} p_2' - \filter p_1 \qc \\
        s_1' &= \filter \delay{2'} p_3 - \filter p_1' \qs
    \end{aligned}
    \label{eq:definition-s1}
\end{equation}
In the absence of secondary noise sources, the expressions for the test mass and reference signals are equal and given by
\begin{equation}
    \begin{aligned}
    \epsilon_1 &= \tau_1 = \filter p_1' - \filter p_1 \qc \\
    \epsilon_1' &= \tau_1' = \filter p_1 - \filter p_1' \qs
    \end{aligned}
    \label{eq:definition-epsilon1}
\end{equation}

\section{Time-delay interferometry}
\label{sec:tdi}

\begin{figure}
    \centering
    \includegraphics[width=\columnwidth]{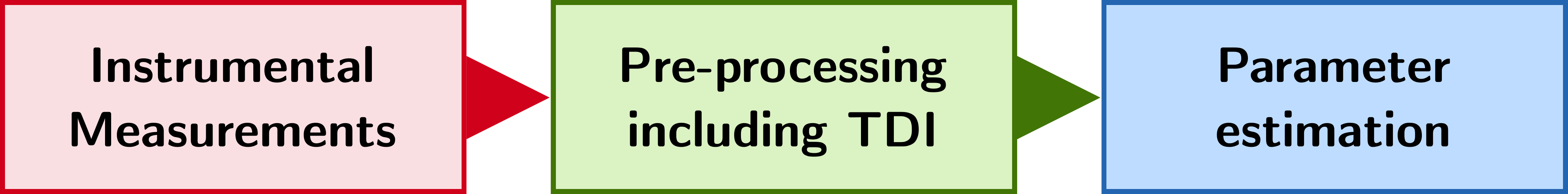}
    \caption{Main steps of the analysis chain.}
    \label{fig:chain}
\end{figure}

\Gls{tdi} is a multistaged algorithm~\cite{Otto:2015wp} which is performed off-line, before astrophysical and cosmological source parameters are extracted (cf.~\cref{fig:chain}). It is mainly conceived to reduce reduce laser frequency noise, but intermediary steps also cancel other instrumental sources of noise. We shall only consider laser frequency noise, but will nevertheless use the full \gls{tdi} expressions in this section, for consistency with the literature.

One first uses the measurement signals to form the intermediary variables $\xi_i$, then $Q_i$ and then finally $\eta_i$. These combinations, respectively, cancel out the optical bench displacement noise (here set to zero), reduce the signals to one free-running laser per spacecraft and suppress clock noise (here set to zero). Note that the test mass acceleration and optical measurement system noises, also set to zero in this study, are not suppressed by these combinations. $\xi$, $Q$, and $\eta$ are defined in~\cite{Otto:2015wp} and can be written under our assumptions as
\begin{equation}
    \begin{gathered}
        \xi_i = s_i - \frac{\delay{k} \epsilon_j' - \delay{k} \tau_j'}{2} \qc \\
	    \xi_i' = s_i' - \frac{\delay{j'} \epsilon_k - \delay{j'} \tau_k}{2} \qc \\
        Q_i = \xi_i + \frac{\delay{k} \tau_j' - \delay{k} \tau_j}{2} \qc \\
        Q_i' = \xi_i' + \frac{\tau_i' - \tau_i}{2} \qc \\
        \eta_i = Q_i \qc \\
        \eta_i' = Q_i' \qs
    \end{gathered}
    \label{eq:definition-intermediary-variables}
\end{equation}

Next, \gls{tdi} synthesizes virtual equal-arm interferometric measurements in order to reduce laser frequency noise. This is done by applying one of the appropriate sets of nested delays to the $\eta_i$ variables, and by combining the resulting terms. In this paper, we focus on the Michelson variables $X, Y$ and $Z$, which synthesize pairwise independent Michelson-like interferometers. There exist several generations of Michelson variables, which depend on the complexity of the spacecraft motion. \Gls{tdi} version 1.0 applies to a static configuration. Version 1.5 applies to a rigid but rotating configuration. Finally, \gls{tdi} version 2.0 applies to a rotating configuration with armlengths varying linearly in time. In this paper we focus on versions 1.5 and 2.0 of \gls{tdi}.

\begin{widetext}
The expressions for the $X$ variable for generations 1.5 and 2.0~\cite{Dhurandhar:2002zcl} are given by
    \begin{align}
        X_1 ={}& \eta_{1'} + \delay{2'} \eta_3 + \delay{2'2} \eta_1 + \delay{2'23} \eta_{2'} - \qty( \eta_1 + \delay{3} \eta_{2'} + \delay{33'} \eta_{1'} + \delay{33'2'} \eta_3) \qc
        \label{eq:X1-wrt-eta}
        \\
        \begin{split}
            X_2 ={}& X_1 + \delay{2'233'} \eta_1 + \delay{2'233'3} \eta_{2'} + \delay{2'233'33'} \eta_{1'} + \delay{2'233'33'2'} \eta_3 \\
            &- (\delay{33'2'2} \eta_{1'} + \delay{33'2'22'} \eta_3 + \delay{33'2'22'2} \eta_1 + \delay{33'2'22'23} \eta_{2'}) \qc
        \end{split}
        \label{eq:X2-wrt-eta}
    \end{align}
where we have used the nested delay notation $\delay{i_1 i_2 \dots i_n} \equiv \delay{i_1} \delay{i_2} \dots \delay{i_n}$. The remaining Michelson variables $Y$ and $Z$ can be obtained by circular permutation of the indices.

Substituting in \cref{eq:X1-wrt-eta,eq:X2-wrt-eta} the values for $\eta$ and $\eta'$, given by \cref{eq:definition-intermediary-variables}, yields expressions which depend on laser frequency noises only. Let us first neglect the filters, \textit{i.e.}~we set $\filter = 1$. Because we are dealing with time-varying armlengths, the Michelson variables can be expressed with nonvanishing delay commutators $\comm{A}{B} = AB-BA$,
\begin{align}
    X_1 ={}& \qty(
        \comm{\delay{2'2}}{\delay{3}} \delay{3'}
        + \delay{3} \comm{\delay{2'2}}{\delay{3'}}
    ) p_1 \qc
    \label{eq:X1-without-filter} \\
    X_2 ={}& \qty(
        \comm{\delay{2'2}}{\delay{33'}} \delay{33'2'2}
        - \delay{33'2'2} \comm{\delay{2'2}}{\delay{33'}}
    ) p_1 \qs
    \label{eq:X2-without-filter}
\end{align}

If we now include the effect of the filter, delay-filter commutators appear and the residual laser frequency noise now reads
\begin{align}
    \begin{split}
        X_1 ={}& \qty(
            \comm{\delay{2'2}}{\delay{3}} \filter\delay{3'}
            + \delay{3} \comm{\delay{2'2}}{\delay{3'}} \filter
        ) p_1 \\
        &+ \delay{3} \qty(
            1
            - \delay{2'2}
        ) \comm{\delay{3'}}{\filter} p_1
        + \qty(
            1
            - \delay{33'}
        ) \delay{2'} \comm{\filter}{\delay{2}} p_{1'} \\
        &+ \qty(
            1
            - \delay{2'2}
        ) \comm{\delay{3}}{\filter} p_{2'}
        + \qty(
            1
            - \delay{33'}
        ) \comm{\delay{2'}}{\filter} p_3 \qc
    \end{split}
    \label{eq:X1-with-filters} \\
    \begin{split}
        X_2 ={}& \qty(
            \comm{\delay{2'2}}{\delay{33'}} \delay{33'2'2}\filter
            + \delay{33'2'2} \comm{\delay{33'}}{\delay{2'2}} \filter
        ) p_1 \\
        &+ \qty(
            1
            - \delay{2'2}
            - \delay{2'233'}
            + \delay{33'2'22'2}
        ) \delay{3} \comm{\delay{3'}}{\filter} p_1 \\
        &+ \qty(
            1
            - \delay{33'}
            - \delay{33'2'2}
            + \delay{2'233'33'}
        ) \delay{2'} \comm{\filter}{\delay{2}} p_{1'} \\
        &+ \qty(
            1
            - \delay{2'2}
            - \delay{2'233'}
            + \delay{33'2'22'2}
        ) \comm{\delay{3}}{\filter} p_{2'} \\
        &+ \qty(
            1
            - \delay{33'}
            - \delay{33'2'2}
            + \delay{2'233'33'}
        ) \comm{\filter}{\delay{2'}} p_3 \qs
    \end{split}
    \label{eq:X2-with-filters}
\end{align}

\end{widetext}

In the next section, we use these expressions to derive an analytic model for the residual laser frequency noise after application of \gls{tdi}. In \cref{sec:simulation}, we numerically simulate the measurement signals, generate the $X$, $Y$, and $Z$ variables, and estimate their \glspl{psd}.

\section{Analytic modeling for linear armlengths}
\label{sec:analytic-model}

For realistic spacecraft orbits computed using Kepler's laws~\cite{Nayak:2006zm,Dhurandhar:2004jr}, \gls{lisa} armlengths are not constant, but modulated with a characteristic time scale of a year. In this section, we expand these armlengths to first order in time. This is a good approximation of the true orbits on a scale of days, \textit{i.e.}~to a frequency of the order of \SI{E-5}{Hz}, while the lowest frequency in \gls{lisa}'s band of interest is \SI{2E-5}{Hz}~\cite{MRDv1}. Therefore, deviations are expected to appear only at low frequencies.

We define the armlengths as $L_i(t) = L_i + \dot{L_i} \, t$, where $L_i$ and $\dot{L_i}$ are constant. The delay operators $\delay{i}$ applied to the laser frequency noise $p(t)$ is now a pure delay and a time rescaling. It reads
\begin{equation}
    \delay{i}p(t) = p \qty[ \qty(1 - \dot{L_i}) t - L_i)] \qs
    \label{eq:linear-delay}
\end{equation}
Its Fourier transform is given by
\begin{equation}
    \frac{1}{1-\dot{L}_i} \exp(-j\omega \frac{L_i}{1-\dot{L}_i}) \fourier{x}(\frac{\omega}{1-\dot{L}_i}) \qc
    \label{eq:linear-delay-tf}
\end{equation}
where $j^2=-1$ is the unit imaginary number.

\subsection{Delay commutators}

\Cref{eq:X1-without-filter,eq:X2-with-filters} are written as sums of delay and delay-filter commutators. In order to compute the \gls{psd} of these expressions, let us first derive their Fourier transforms.

The expression of nested delay operators in the time domain is deduced by repeated use of \cref{eq:definition-delay-operator}. One finds
\begin{equation}
    \delay{i_1 \dots i_n} x(t) = x\qty(S_n t - \sum_{k=1}^n{S_{k-1} L_{i_k}}) \qc
    \label{eq:nested-delays}
\end{equation}
where we have defined the product $S_k = \prod_{p=1}^k{(1-\dot{L}_{i_p})}$ for $k>0$, and $S_0=1$.
The Fourier transform of \cref{eq:nested-delays} is given by
\begin{equation}
    \frac{1}{S_n} \exp(-j\omega \sum_{k=1}^n{\frac{L_{i_k}}{S_k}}) \fourier{x}(\frac{\omega}{S_n}) \qs
    \label{eq:nested-delays-tf}
\end{equation}

Let us now consider the commutator of $n$ delay operators applied to a signal $x(t)$, which we denote
$y(t) = \comm{\delay{i_1 \dots i_m}}{\delay{i_{m+1} \dots i_n}} x(t) = \delay{i_1} \dots \delay{i_n} x(t) - \delay{i_{m+1}} \dots \delay{i_n} \delay{i_1} \dots \delay{i_m} x(t)$ in the following. Using \cref{eq:nested-delays} and after some work on the indices, it is possible to express it as
\begin{equation}
    \begin{split}
        & y(t) = x\qty(S_n t - \sum_{k=1}^n{S^i_{k-1} L_{i_k}})
        \\
        & \quad - x \qty(S_n t - \frac{S^i_n}{S^i_m} \sum_{k=1}^{m}{S^i_{k-1} L_{i_k}} - \frac{1}{S^i_m} \sum_{k=m+1}^{n}{S^i_{k-1} L_{i_k}}) \qs
    \end{split}
    \label{eq:delay-commutator}
\end{equation}

We can expand this expression to first order in powers of the armlength derivatives $\dot{L}_i$. If we moreover assume that all armlengths at $t=0$ are almost equal, \textit{i.e.}~$L_i \approx L$ for all $i$, it reads
\begin{equation}
    y(t) \approx L \qty[ (n-m) \qty(\sum_{k=1}^{m}{\dot{L}_{i_k}}) - m \qty(\sum_{k=m+1}^{n}{\dot{L}_{i_k}})] \, \dv{x}{t} (t-nL) \qs
    \label{eq:delay-commutator-approx}
\end{equation}
The corresponding Fourier transform $\fourier{y}(\omega)$ is
\begin{equation}
    \begin{split}
        & \fourier{y}(\omega) \approx -j\omega L e^{-j\omega nL}
        \\
        & \quad \times \qty[\qty(n-m) \qty(\sum_{k=1}^{m}{\dot{L}_{i_k}}) - m \qty(\sum_{k=m+1}^{n}{\dot{L}_{i_k}})] \fourier{x}(\omega) \qs
    \end{split}
    \label{eq:delay-commutator-ft}
\end{equation}

\subsection{Delay-filter commutators}

In the time domain, the application of the filter $\filter$ on a signal $x(t)$ is written as the convolution of the latter and the filter's kernel $f$.  In frequency domain, this translates into the product $\fourier{x}(\omega) \fourier{f}(\omega)$.

Therefore, if we apply the filter after a series of delays $\filter \delay{i_1} \dots \delay{i_n} x(t)$, we have, in Fourier domain,
\begin{equation}
    \frac{1}{S_n} \exp(-j\omega \sum_{k=1}^n{\frac{L_{i_k}}{S_k}}) \fourier{x}(\frac{\omega}{S_n}) \fourier{f}(\omega) \qs
\end{equation}
If the filter is applied before the delays $\delay{i_1} \dots \delay{i_n} \filter x(t)$, we now have
\begin{equation}
    \frac{1}{S_n} \exp(-j\omega \sum_{k=1}^n{\frac{L_{i_k}}{S_k}}) \fourier{x}(\frac{\omega}{S_n}) \fourier{f}(\frac{\omega}{S_n}) \qs
\end{equation}

Let us define the signal $y(t)$ as the commutator of nested delays $\delay{i_1} \dots \delay{i_n}$ and a filter $\filter$. In general,
\begin{equation}
    \begin{split}
        y(t) &= \comm{\delay{i_1} \dots \delay{i_n}}{\filter} x(t)
        \\
        &= \delay{i_1} \dots \delay{i_n} \filter x(t) - \filter \delay{i_1} \dots \delay{i_n} x(t) \qs
    \end{split}
    \label{eq:delay-filter-commutator}
\end{equation}
Using the previous equations, the exact expression in Fourier space writes
\begin{equation}
    \begin{split}
        \fourier{y}(\omega) &= \frac{1}{S_n} \exp(-j\omega \sum_{k=1}^n{\frac{L_{i_k}}{S_k}})
        \\
        & \qquad \times \fourier{x}(\frac{\omega}{S_n}) \qty[ \fourier{f}(\frac{\omega}{S_n}) - \fourier{f}(\omega) ] \qs
    \end{split}
    \label{eq:delay-filter-commutator-tf}
\end{equation}

If we use a first-order expansion in the armlength derivatives $\dot{L}_i$, the previous equation reads
\begin{equation}
    \fourier{y}(\omega) \approx \omega \exp(-j\omega \sum_{k=1}^n{L_{i_k}}) \qty(\sum_{k=1}^n{\dot{L}_{i_k}}) \dv{\fourier{f}}{\omega} (\omega) \,\fourier{x}(\omega) \qs
    \label{eq:delay-filter-commutator-approx-tf}
\end{equation}
One can note the linear dependency on the angular frequency, and the first-order factor $\sum_{k=1}^n{\dot{L}_{i_k}}$. The term of interest here is $\dv{\fourier{f}}{\omega} (\omega)$, which depends on the filter characteristics.

\subsection{Residual laser noise}

First, let us neglect the filters; \textit{i.e.}, we set $\filter = 1$. We substitute in \cref{eq:X1-without-filter} the first-order expression for the delay commutator given in \cref{eq:delay-commutator-ft}. This yields the approximated Fourier transform of the residual laser noise for $X_1$.

As expected, first-order terms vanish for \gls{tdi} 2.0. We expand \cref{eq:X2-without-filter} to second order, using the exact expression of the delay commutator given in \cref{eq:delay-commutator-ft}. This yields the approximated Fourier transform of the residual laser noise for $X_2$.

The corresponding \glspl{psd} are obtained by taking the squared modulus and the ensemble average of the Fourier transforms. We use the fact that the laser noises have zero mean, \textit{i.e.}~$\expval{\fourier{p_i}(\omega)} = 0$ for all $i$. In addition, different laser noises are uncorrelated, \textit{i.e.}~$\expval{\fourier{p_i}(\omega_1)\fourier{p_j}(\omega_2)} = 0$ if $i \neq j$. They all are white noises with the same constant \gls{psd}, denoted $\psd{p}$. We have
\begin{align}
    \psd{X_1}(\omega) &\approx
	16 \psd{p} \, \omega^2 L^2 \qty(\dot{L_2} - \dot{L_3})^2 \qc
	\label{eq:X1-residuals-without-filter} \\
	\psd{X_2}(\omega) &\approx
	64 \psd{p} \, \omega^2 L^2 \qty(\dot{L_2}^2 - \dot{L_3}^2)^2 \qs
	\label{eq:X2-residuals-without-filter}
\end{align}
As expected, the residual laser noise scales with the laser frequency noise $\psd{p}$, and vanishes if $\dot{L}_1 = \dot{L}_2$, \textit{i.e.}~if the constellation undergoes a homothetic transformation.

We now introduce the effect of the filters. Using \cref{eq:X1-with-filters,eq:delay-filter-commutator-approx-tf}, we find that the first-order expansion of the laser noise residuals in $X_1$ is given by
\begin{widetext}
\begin{equation}
    \psd{X_1} \approx 8 \psd{p} \omega^2 \qty[2 L^2 \qty(\dot{L}_2 - \dot{L}_3)^2 \psd{f}(\omega) - L \dot{L}_3 \qty(\dot{L}_2 - \dot{L}_3) D_\filter(\omega) + \sin[2](\omega L) \qty(\dot{L}_2^2 + \dot{L}_3^2) K_\filter(\omega)] \qc
\end{equation}
\end{widetext}
where we have defined the squared modulus of the filter transfer function $\psd{f}(\omega) = \abs{\fourier{f}(\omega)}^2$, the filter term $K_\filter(\omega) = \abs{\dv{\fourier{f}}{\omega} (\omega)}^2$, and $D_\filter(\omega) = \Im{\qty(1 - e^{-j2L \omega}) \fourier{f}(\omega) \dv{\fourier{f}^*}{\omega}} = \qty[1 - \cos(2L\omega)] \Im{\fourier{f}(\omega) \dv{\fourier{f}^*}{\omega}} - \sin(2L\omega) \Re{\fourier{f}(\omega) \dv{\fourier{f}^*}{\omega}}$. Comparing this expression with \cref{eq:X1-residuals-without-filter}, we see that an extra term of the same order appears. It corresponds to a coupling between the antialiasing filters and the time-varying armlengths, with a dependance on the filter characteristics expressed by the filter term $K_\filter(\omega)$ and $D_\filter(\omega)$, discussed below. This flexing-filtering coupling is however smaller than the previous term by a factor of $1/L$, and \cref{eq:X1-residuals-without-filter} still gives a good estimate for the residual laser noise in $X_1$.

Similarly, we use \cref{eq:delay-filter-commutator-approx-tf} in \cref{eq:X2-with-filters} to obtain the approximated expression for the laser noise residuals in $X_2$, which reads
\begin{equation}
    \begin{split}
        \psd{X_2}(\omega) &\approx
        32 \psd{p} \omega^2
        \sin[2](\omega L) \sin[2](2\omega L)
        \\
        &\qquad \times \qty(\dot{L_2}^2 + \dot{L_3}^2)
        K_\filter(\omega) \qs
    \end{split}
    \label{eq:X2-residuals-with-filter}
\end{equation}
We see that the flexing-filtering coupling is dominant for second-generation \gls{tdi}. The level of residual laser frequency noise in $X_2$ is therefore strongly dependent on the filter design. We study various filters in the next paragraphs.

\subsection{Filter term \texorpdfstring{$K_\filter(\omega)$}{Kf}}

In the current baseline, all antialiasing filters are identical and correspond to a causal symmetrical \gls{fir} filter. Its corner frequencies are slightly below \SI{1}{Hz}, and a high attenuation must be reached for frequencies higher than \SI{2}{Hz}.

We can write the filter output $y_n$ as a function of the past input samples $x_{n-k}$ and $2N+1$ coefficients $\alpha_k$
\begin{equation}
    y_n = \sum_{k=0}^{2N}{\alpha_k x_{n-k}} \qs
\end{equation}
Its transfer function reads
\begin{equation}
    \fourier{f}(\omega) = \sum_{k=0}^{2N}{\alpha_k e^{-jk \omega/f_s}} \qc
    \label{eq:fir-transfer-function}
\end{equation}
$f_s$ is the sampling frequency. Taking its derivative with respect to the angular frequency $\omega$ immediately yields the associated filter term
\begin{equation}
    K_\filter^\text{causal}(\omega) = f_s^{-2} \abs{
        \sum_{k=1}^{2N}{k \alpha_k e^{-j k \omega /f_s}}
    }^2 \qs
    \label{eq:filter-term-causal}
\end{equation}

This causal filter has a nonvanishing group delay of $N f_s^{-1}$, which is responsible for the nonvanishing zeroth-order term $K_\filter^\text{causal}(\omega) \approx f_s^{-2} \abs{\sum_{k=1}^{2N}{k\alpha_k}}^2$, for $\omega \ll 2 \pi f_s$. The equivalent noncausal filter has a vanishing group delay, and the associated filter term reads
\begin{equation}
    K_\filter^\text{acausal}(\omega) = 4 f_s^{-2} \abs{
        \sum_{k=1}^{N}{k \alpha_{N+k} \sin(\frac{k \omega}{f_s})}
    }^2 \qs
    \label{eq:filter-term-approx-acausal}
\end{equation}
We are now left with a second-order term in $\omega / 2\pi f_s$, and $K_\filter^\text{acausal}(\omega) \approx 4 \omega^2 f_s^{-4} \abs{\sum_{k=1}^{N}{k^2 \alpha_{N+k}}}^2$.

\begin{figure*}
    \centering
    \includegraphics[width=\textwidth]{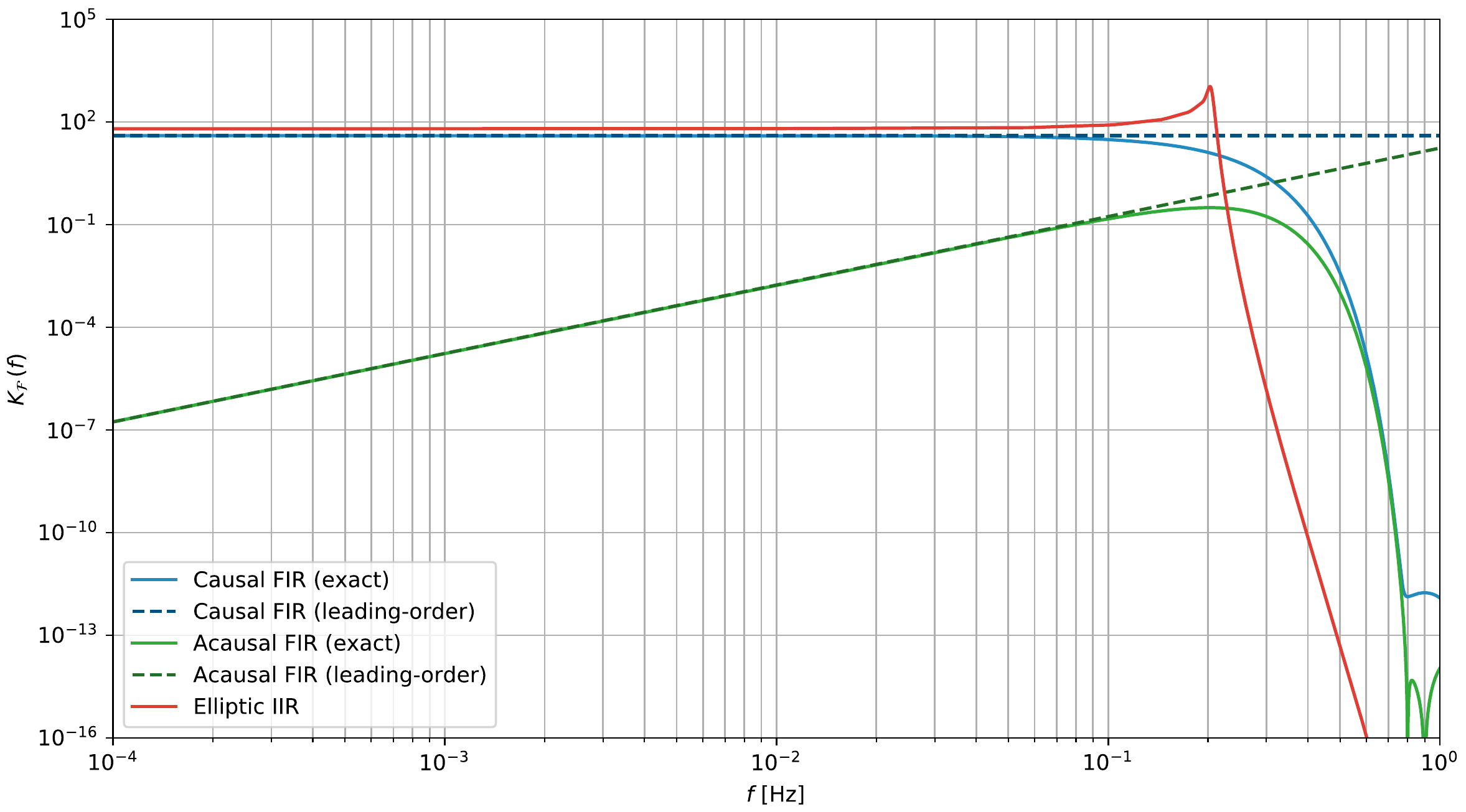}
    \caption{Levels for different filter terms $K_\filter$. Dotted lines correspond to leading-order expansions.}
    \label{fig:filter-terms}
\end{figure*}

We expect smaller laser noise residuals for noncausal filters in the \gls{lisa} frequency band of interest, \textit{i.e.}~below \SI{1}{Hz}. This is verified in \cref{fig:filter-terms}, where causal and noncausal filter terms are plotted, for the filter used in simulations and described in \cref{sec:simulation}.

For reference, we also plotted the filter term for an \gls{iir} (\textit{i.e.}~recursive) elliptic filter with the same characteristics\footnote{Coefficients of the filter are given in \cref{app:elliptic-coefficients}.}. We see that it is larger than that of the noncausal \gls{fir} filters, which leads to larger residual noise.

\section{Simulations for Keplerian orbits}
\label{sec:simulation}

In this section, we present \texttt{LISACode} and \texttt{LISANode}, the two simulation softwares that were used to generate \gls{lisa} measurement signals and process them using the \gls{tdi} algorithm introduced in \cref{sec:tdi}. The results are presented and discussed in \cref{sec:results}.

\texttt{LISACode} is the simulator currently used by the \gls{lisa} community to generate realistic data while \texttt{LISANode} is the baseline prototype for an end-to-end mission performance simulator. They both perform computations in time domain and produce time series for any choice of \gls{tdi} variables. In the following, we only consider the Michelson variables $X$, $Y$, and $Z$.

Two sampling frequencies are used in our simulations. The \textit{physical} sampling frequency applies to the physical subsystems in the simulators: generation of instrumental noise, beam propagation, and optical measurements. It is taken to be equal to $f_\text{phy} = \SI{20}{Hz}$ in both simulators. The interferometric signals $s_i$, $\epsilon_i$ and $\tau_i$ are downsampled to the \textit{measurement} frequency $f_\text{meas} = \SI{2}{Hz}$ by means of a decimation algorithm. All preprocessing steps, including \gls{tdi}, are therefore carried out at this measurement frequency. All signals are implemented as doubles (64-bit floating-point numbers).

In both simulators, we use a symmetric \gls{fir} antialiasing filter of order 253, designed with a Kaiser window. The coefficients\footnote{Given in \cref{app:fir-coefficients}.} are calculated such that the signal is attenuated by \SI{240}{dB} between \SI{0.2}{Hz} and \SI{0.9}{Hz}, and we authorize a maximum ripple of \SI{0.1}{dB} below \SI{0.2}{Hz}. We implement filters using a direct form~I and therefore, account group delays when they are not vanishing.

The propagation of the laser beams between the spacecraft is implemented using time-varying delays. Those delays are computed from the relative positions of the spacecraft, themselves deduced from their Keplerian orbits presented in~\cite{Nayak:2006zm}. These orbits include the Sagnac effect, as well as first order relativistic corrections.

All delay operators are implemented using Lagrange interpolating polynomials of order 31. This choice is the result of a trade-off: it allows for good precision and limits execution time and numerical errors. As seen above, the \gls{tdi} algorithm requires the application of multiple delay operators to the interference measurements for the calculation of the Michelson variables $X$, $Y$, and $Z$. In order to minimize the error introduced by the associated interpolations, we use a nested delay algorithm in which a single interpolation is necessary.

\subsection{LISACode}

\texttt{LISACode} is a high-level simulator~\cite{Petiteau:2008wa}, entirely written in \texttt{C++}. It was used to produce noise time series for the past \glspl{mldc}~\cite{Babak:2008kx} as well as for the current \gls{ldc}. It also constitutes a useful tool for the various studies of the instrument noise budget.

\begin{figure}
    \centering
    \includegraphics[width=\columnwidth]{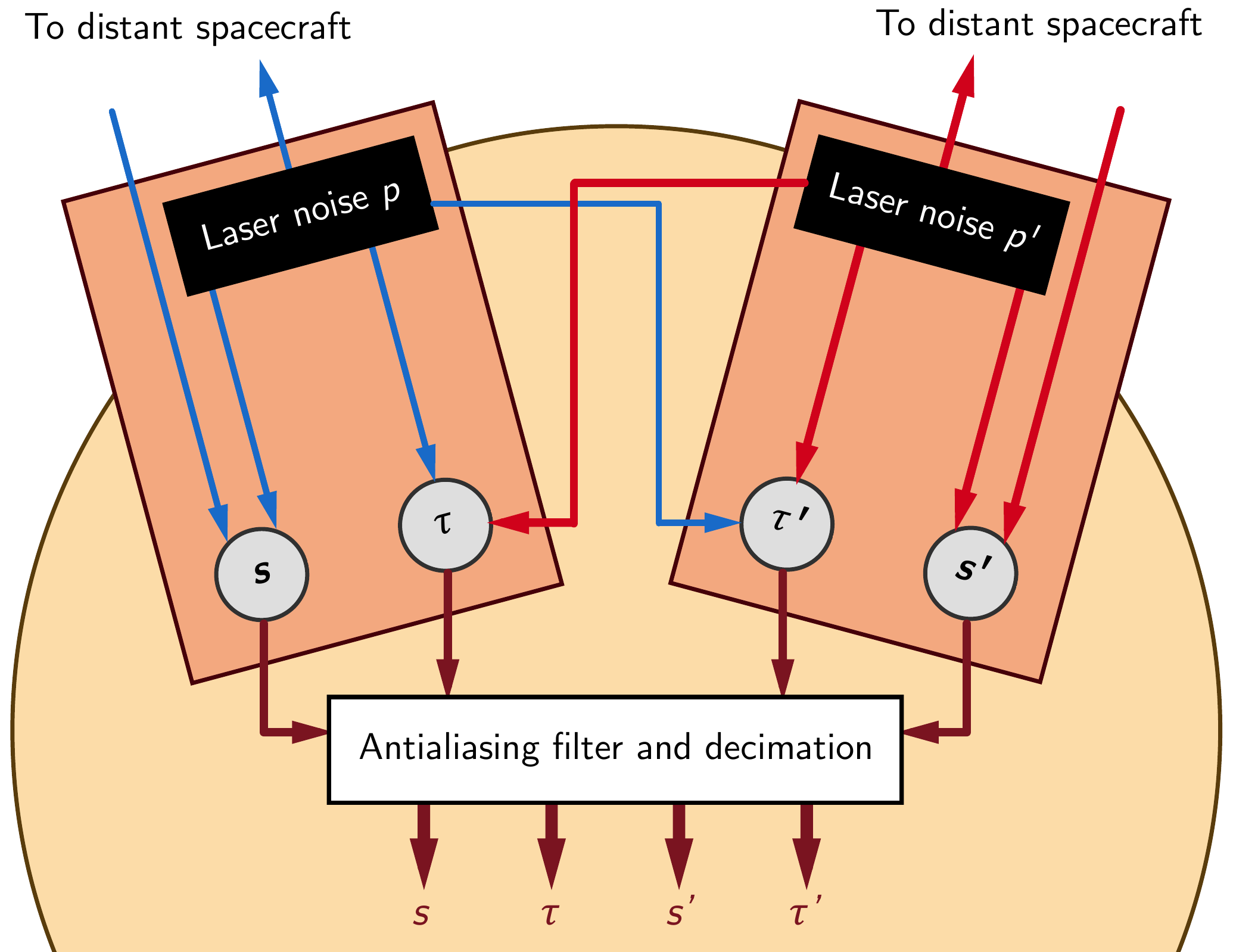}
    \caption{Original optical design used in \texttt{LISACode} simulations. Four interferometric measurements per spacecraft are performed: the science signals $s_i$ and $s_i'$, along with the reference signals $\tau_i$ and $\tau_i'$.}
    \label{fig:lisacode}
\end{figure}

\texttt{LISACode} is based on the original optical design, equivalent, under our assumptions, to the \textit{split interferometry} design described in \cref{sec:instrumental-setup}. Each of the three spacecraft of the \gls{lisa} constellation contains two independent lasers and two optical benches. Each optical bench holds a science and a reference interferometer; the corresponding beat notes are filtered and decimated to produce the respective measurement signals $s_i(t)$ and $\tau_i(t)$, as presented in \cref{fig:lisacode}.

The \texttt{LISACode} results use a \SI{E7}{s} time series generated with version 2.12.

\subsection{LISANode}

\texttt{LISANode}~\cite{Bayle:tbp} is a flexible simulation tool based on the foundations of \texttt{LISACode}, which aims to assess \gls{lisa}'s scientific performance. It is the current prototype for an end-to-end simulator of the mission. It was originally developed by the authors and is now part of the \gls{lisa} Simulation Group activities.

Similarly to \texttt{LISACode}, \texttt{LISANode} works exclusively in the time domain so that it can handle nonlinear artifacts and produce output in the form of time series. It is based on simulation graphs, written in \texttt{Python}, which are composed of atomic \texttt{C++} computation units called nodes. A scheduler triggers node execution in a specific order and pushes data from one node to the next. In this manner, execution time is optimized and data flow is synchronized. Graphs can be nested to represent whole subsystems as one object, allowing for a high level of modularity and maintainability.

\begin{figure}
    \centering
    \includegraphics[width=\columnwidth]{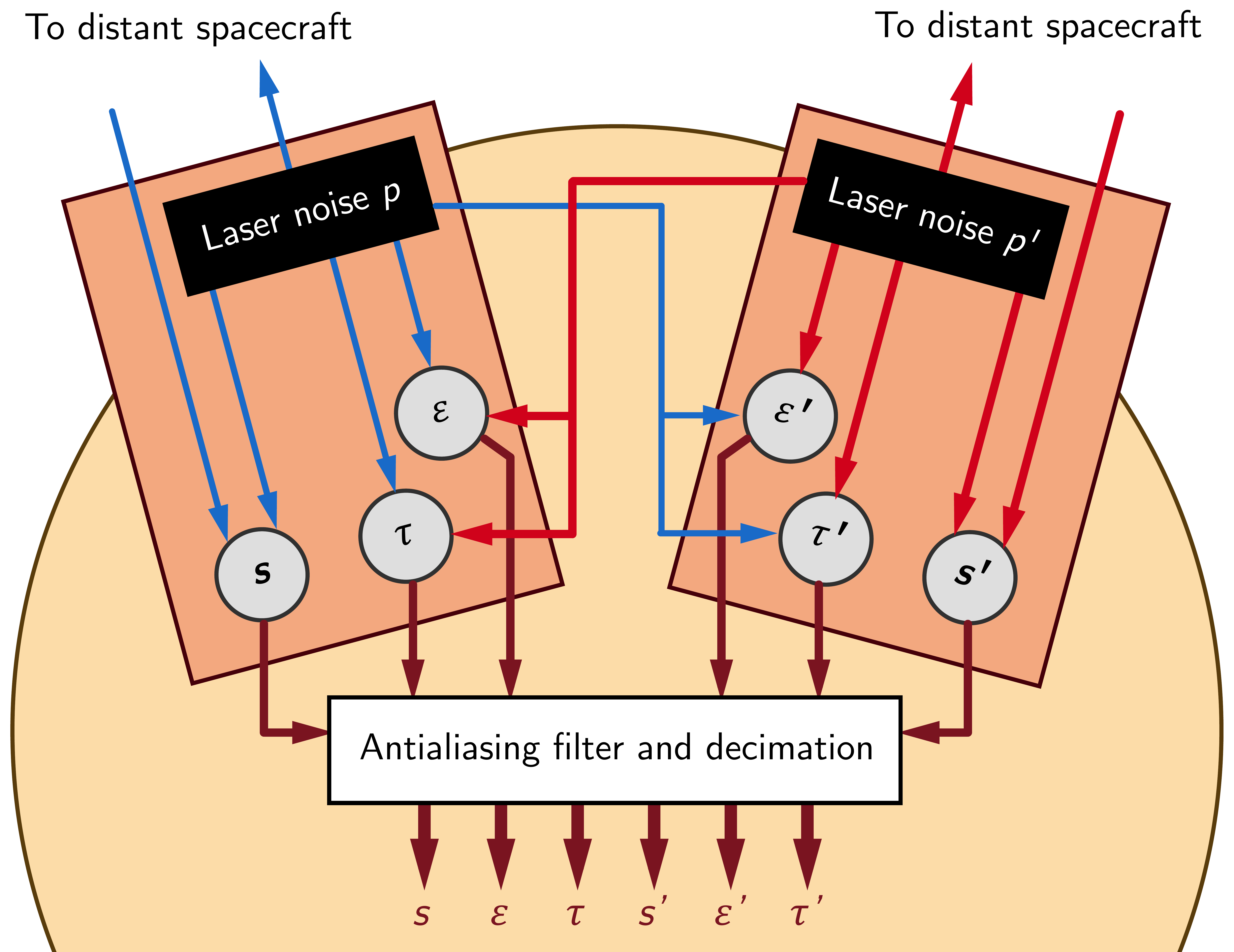}
    \caption{New \textit{split interferometry} optical design used in \texttt{LISANode} simulations. Six interferometric measurements per spacecraft are performed: the science signals $s_i$ and $s_i'$, the test mass signals $\epsilon_i$ and $\epsilon_i'$, along with the reference signals $\tau_i$ and $\tau_i'$.}
    \label{fig:lisanode}
\end{figure}

\texttt{LISANode} implements the newest \textit{split interferometry} optical setup described in \cref{sec:instrumental-setup}, and presented in \cref{fig:lisanode}. Three interferometric measurements $s_i$, $\tau_i$ and $\epsilon_i$ (respectively, the science, test mass, and reference signals) are formed and relevant sources of noise are added to the measurements. These signals are then transmitted to the on-board computer, which contains the antialiasing filter and decimation nodes. The results of these operations are used to form the \gls{tdi} Michelson variables $X$, $Y$ and $Z$.

The results use a \SI{E7}{s} time series generated with \texttt{LISANode} version 1.1.

\section{Results and discussion}
\label{sec:results}

\subsection{Results}
\label{sub:results}

\begin{figure*}
    \centering
    \includegraphics[width=\textwidth]{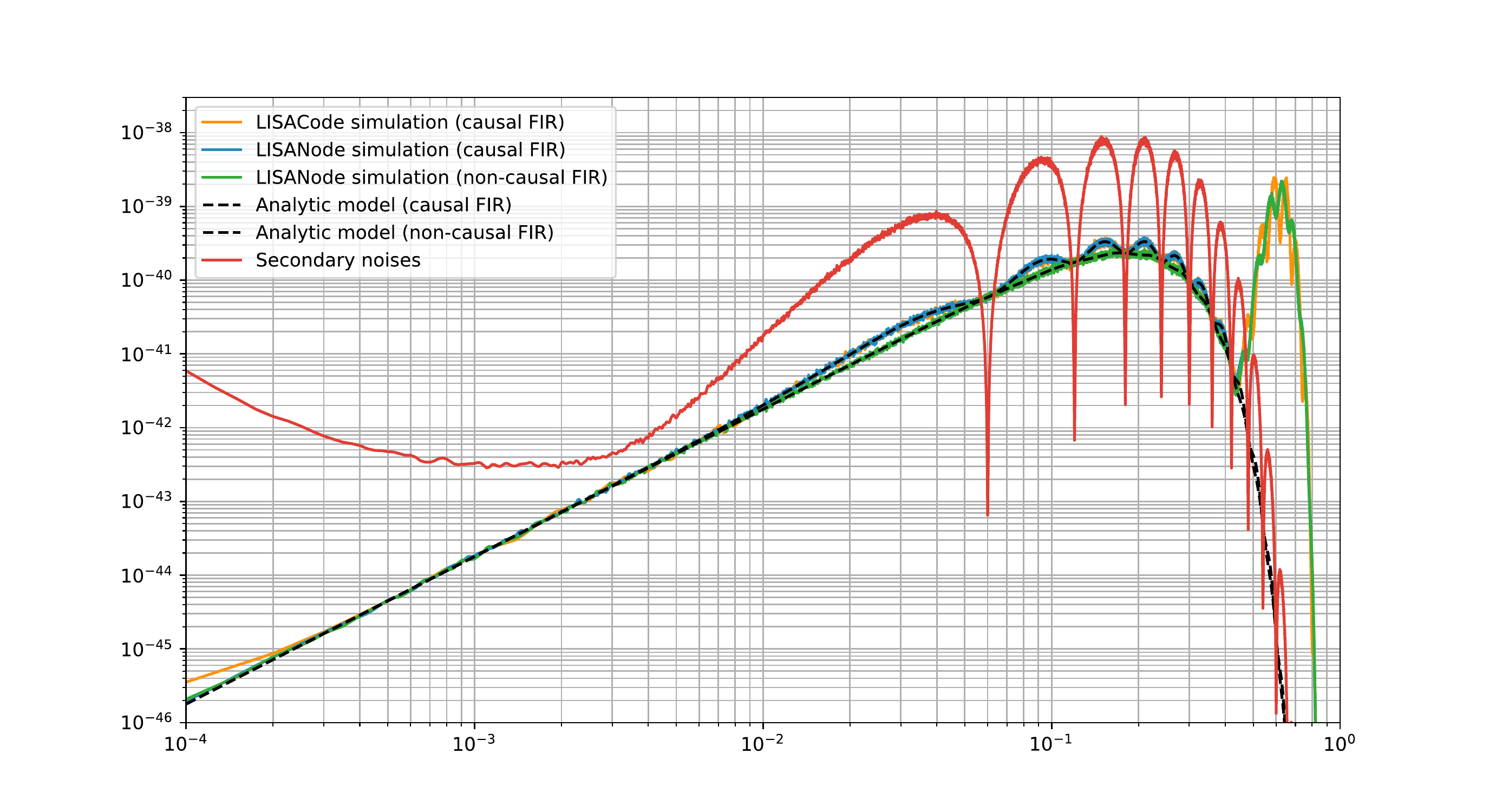}
    \caption{Power spectral density of the residual laser frequency noise in the Michelson $X_1$ channel. The \texttt{LISACode} and \texttt{LISANode} simulations use realistic Keplerian orbits, while the theoretical model uses armlengths varying linearly with time. Secondary noises are shown in red and indicate the target level of laser frequency noise suppression.}
    \label{fig:tdi1}
\end{figure*}

\begin{figure*}
    \centering
    \includegraphics[width=\textwidth]{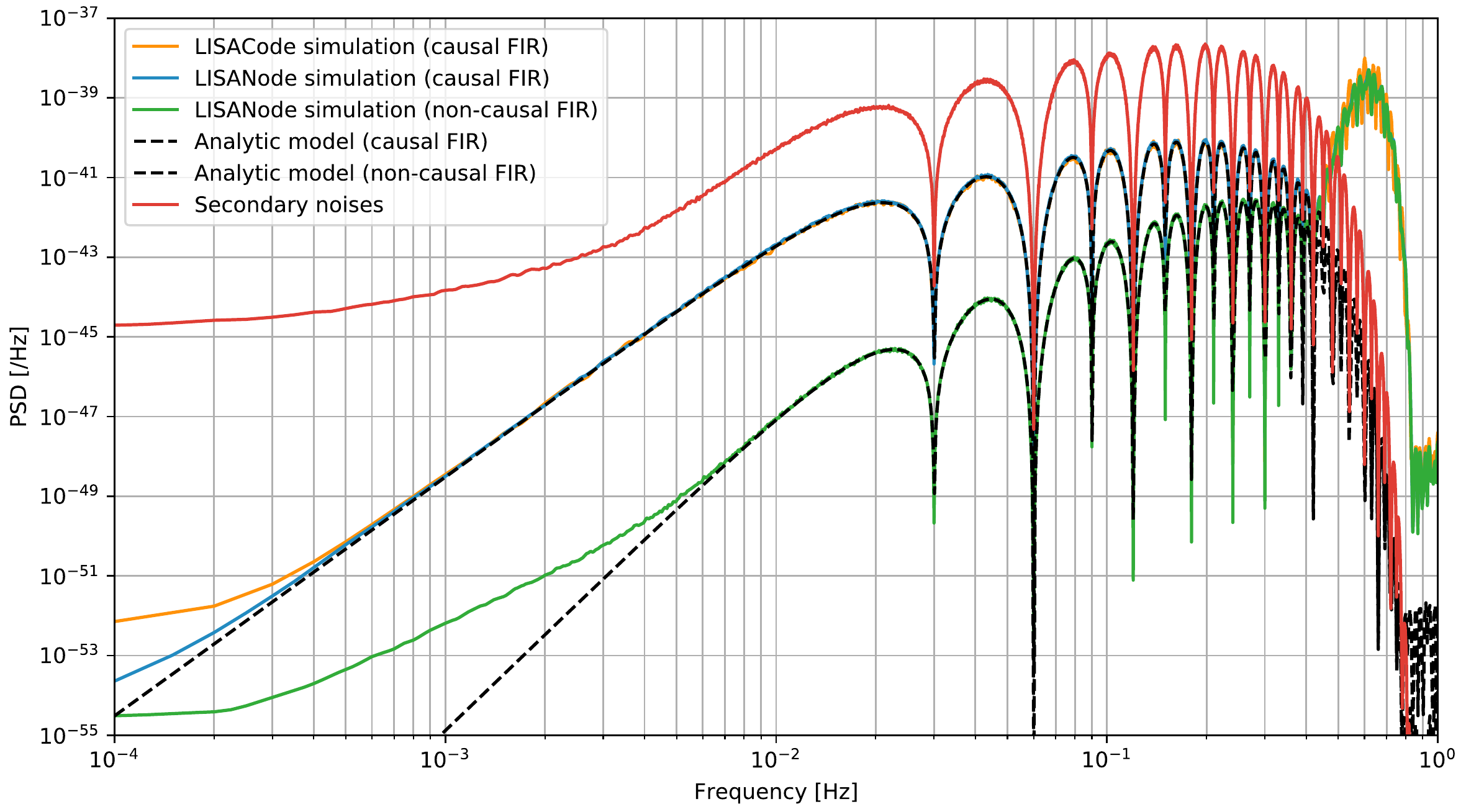}
    \caption{Power spectral density of the residual laser frequency noise in the Michelson $X_2$ channel. The \texttt{LISACode} and \texttt{LISANode} simulations use realistic Keplerian orbits, while the theoretical model uses armlengths varying linearly with time. Secondary noises are shown in red and indicate the target level of laser frequency noise suppression.}
    \label{fig:tdi2}
\end{figure*}

In \cref{fig:tdi1,fig:tdi2}, we present the \glspl{psd} of the residual laser frequency noise for the \gls{tdi} Michelson variables $X_1$ and $X_2$. We show the results of \texttt{LISANode} simulations for both the causal and the noncausal versions of the same filter, as described in \cref{sec:simulation} (light and dark blue curves). We plot the results of \texttt{LISACode} simulations for the causal filter only, in order to validate the new simulator (light orange curve). The models derived in \cref{sec:analytic-model} are superimposed (dashed light and dark green curves).

For reference, the red solid curves show the residual secondary noises in both $X_1$ and $X_2$ channels, simulated using \texttt{LISANode} and the noncausal antialiasing filter. To generate those signals, we did not change the simulation parameters. However, laser frequency noise is set to zero while the test mass acceleration (TM), optical read-out (RO) and optical path (OP) noise amplitudes were given their nominal \gls{lisa} instrument noise budget values. The spectral shapes of these three secondary sources of noise are given in~\cite{MRDv1}, and read
\begin{align*}
\psd{\text{TM}} ={}& \qty(\num{2.4E-15})^2 \\
& \times \qty[ 1 + \qty(\frac{\num{4E-4}}{f})^2 ] \si{m^2 s^{-4} Hz^{-1}} \qc \\
\psd{\text{RO}} +{}& \psd{\text{OP}} = \SI{1E-24}{m^2 Hz^{-1}} \qs
\end{align*}
Because \gls{tdi} does not suppress those secondary noises, but only modulates their spectra, they are used as a benchmark.

We use Welch's method to estimate the spectra, implemented with standard \texttt{Python} tools included in the \texttt{scipy.signal} module, version 1.1.0. We use segments of \num{40000} samples and a \texttt{Nutall4} window function. The results are presented for the frequency band from \SI{E-4}{Hz} to \SI{1}{Hz}.

We can see that the results of both simulators are in very good agreement. The fact that both simulators give similar results, although they use different implementations, increases our confidence in the results they produce. Note that at frequencies greater than \SI{4E-1}{Hz}, one observes a slight discrepancy between \texttt{LISANode} and \texttt{LISACode}. This discrepancy is due to different implementations of the Lagrange interpolating polynomials in the two simulators.

We can also observe that our analytic models match the simulated data with exquisite precision in most of the \gls{lisa} band. At high frequencies, the model is no longer valid, since it does not include the errors from Lagrange interpolations. These errors, visible in the simulated data, manifest themselves by an increased level of residual laser frequency noise around \SI{6E-1}{Hz}. It can be shown that varying the interpolation order changes the amplitude of this effect. At lower frequencies, the simulated data deviate away from the analytic model. This is because assuming that the armlengths are varying linearly in time is only valid for frequencies higher than \SI{1}{mHz}. However, we see that at these lower frequencies the residual laser noise is in any case well below mission noise level requirements.

It is also very clear that using a noncausal filter decreases significantly the residual laser noise. This effect is particularly obvious at low frequencies, as the leading-order expansion of the filter term is constant for the causal filter, while being proportional to $\omega^2$ for its noncausal version; see \cref{sec:analytic-model}.

\Cref{fig:tdi1} shows that first-generation Michelson variables reduce the laser frequency noise down to the required level, but only marginally. This is especially true with causal filters. On the contrary, \cref{fig:tdi2} shows that second-generation Michelson variables can reduce the laser frequency noise up to 3 orders of magnitude below the secondary noises over the entire frequency range, if we use noncausal filters.

In the case of time varying orbits, and in the presence of antialiasing filters, \gls{tdi} 2.0 is therefore necessary and sufficient to suppress laser frequency noise levels down to mission requirements~\cite{Audley:2017drz}. Moreover, using noncausal filters allows for comfortable margins.

\subsection{Filter group delay}

\Gls{tdi} uses as inputs the interferometric signals from the six optical benches [$s_i(t)$, $\tau_i(t)$ and $\epsilon_i(t)$], and the ranging estimates $L_i(t)$ for all six links. A causal filter has a nonvanishing group delay $N/f_s$; since it is only applied on the interferometric signals, the latter will be time-shifted while ranging estimates are left untouched. Let us denote the filter group delay operator $\delay{\filter}$. As part of the \gls{tdi} algorithm, one computes terms of the type $\delay{i}\delay{\filter}p(t) = p(t - N/f_s - L_i(t))$ when one really wants $\delay{\filter}\delay{i}p(t) = p(t - N/f_s - L_i(t - N/f_s))$, or, equivalently, $p(t - L_i(t))$.

We recognize here an extra noise proportional to the commutator $\comm{\delay{\filter}}{\delay{i}}p(t) \approx \dot{L}_i (N/f_s) \dv{p}{t} (t - L_i - N/f_s)$; see \cref{eq:delay-commutator-approx}. It is non vanishing in the case of time-varying armlengths, which explains this flexing-filtering coupling.

\subsection{Implementation of noncausal filters}

Using \cref{eq:fir-transfer-function}, we can relate the transfer function $\fourier{f}_\text{causal}$ of the causal version of a filter, to that of the noncausal version $\fourier{f}_\text{acausal}$,
\begin{equation*}
    \fourier{f}_\text{acausal}(\omega) = e^{j\omega N/f_s} \fourier{f}_\text{causal}(\omega) \qc
\end{equation*}
and see that the two only differ by a constant delay of an integer number of samples. This delay exactly matches the group delay of the filter. We can therefore deduce output samples of the noncausal filter by simply retimestamping output samples of the causal version such that
\begin{equation}
    y_k^\text{acausal} = y_{N+k}^\text{causal} \qs
    \label{eq:output-acausal-causal}
\end{equation}

This \textit{relabeling} can be performed by the on-board computer, before data are decimated and telemetered to Earth.

One could instead relabel the telemetered interferometric data off-line. This delay is equal to the filter group delay and might not be an integer number of samples anymore due to the downsampling. Therefore, new interpolation errors enter the measurements. Designing the filter such that its group delay is a multiple of the decimated sampling period could solve this issue. This constraint should be accounted for when designing the on-board software.

In the simulations presented in \cref{sec:simulation,sec:results}, we used the latter implementation.

\section{Conclusion}
\label{sec:conclusion}

This article addresses the problem of modeling and simulating the residual laser frequency noise, after \gls{tdi} has been applied, in a realistic instrumental setup. We have focused our analysis on the first and second-generation Michelson $X$, $Y$ and $Z$ variables, and have included the effect of time-varying armlengths, as well as the effects of the on-board antialiasing filters. In our \texttt{LISACode} and \texttt{LISANode} simulations, the armlengths vary according to Keplerian orbits. In the analytic expressions of the residual laser frequency noise spectrum we derive, armlengths vary linearly with time. This is a very good approximation of Keplerian orbits on the time scales of interest. The resulting expressions are functions of both the varying armlengths and of the filter coefficients, and show at leading-order that a new flexing-filtering coupling noise enters the measurements, degrading the expected \gls{tdi} performance.

We showed that the simulated data match the analytic model with exquisite precision over a large fraction of the \gls{lisa} frequency band. As a benchmark for the performance of \gls{tdi}, we used \texttt{LISANode} simulations that include secondary noise only. In the case of time-varying armlengths, \gls{tdi} 1.5 is shown not be able to achieve sufficient laser frequency noise reduction over the entire frequency range of interest. On the other hand, \gls{tdi} 2.0 reduces laser frequency noise to well below the secondary noise level, for the case of a standard finite-impulse response filter. \Gls{tdi} 2.0 is therefore the minimal viable \gls{tdi} generation for \gls{lisa}.

As demonstrated in this paper, our analytic model and simulations help gain insight into \gls{tdi} and the various parameters that play a key role in its performance. In particular, we were able to demonstrate that a noncausal filter improves \gls{tdi} performance and helps reduce further the residual laser noise down to 3 extra orders of magnitude in the middle of \gls{lisa} frequency band. This noncausal filter can be synthesized using its causal version on board and adapt the \gls{tdi} algorithm by time-shifting the interferometric signals with respect to the ranging estimates. This concept was demonstrated by our simulations.

One could also use the analytic model developed in this study to optimize the filter coefficients, so that the useful frequency band for data analysis (\textit{i.e.}, the frequency band over which the gravitational signal is not attenuated) is maximized, while the residual laser frequency noise level remains below the secondary noise level. Finally, the effect of other instrumental imperfections and artifacts, such as the errors in the absolute ranging or in the interpolation scheme, or even clock noises, remain to be included in both our model and simulations.

\appendix
\section{Elliptic filter coefficients}
\label{app:elliptic-coefficients}

We give the coefficients of the IIR elliptic filter used as a reference in Fig.~\ref{fig:filter-terms}. For numerical stability, the filter is implemented as a series of recursive second-order cells of $z$ transform
\begin{equation*}
    \fourier{f}_\text{cell}(z) = \frac{\alpha_0 + \alpha_1 z^{-1} + \alpha_2 z^{-2}}{1 + \beta_1 z^{-1} + \beta_2 z^{-2}} \qc
\end{equation*}
and a scaling factor of \num{2.587527E-4} applied at the end of the chain.

{\scriptsize First cell: $\alpha =$ \texttt{4.882e-04}, \texttt{7.239e-05}, \texttt{4.882e-04}, $\beta =$ \texttt{-1.969}, \texttt{9.697e-01}. Second cell: $\alpha =$ \texttt{9.765e-04}, \texttt{-1.487e-03}, \texttt{9.765e-04}, $\beta =$ \texttt{-1.971}, \texttt{9.719e-01}. Third cell: $\alpha =$ \texttt{3.125e-02}, \texttt{-5.997e-02}, \texttt{3.125e-02}, $\beta =$ \texttt{-1.992}, \texttt{9.961e-01}. Fourth cell: $\alpha =$ \texttt{7.812e-03}, \texttt{-1.402e-02}, \texttt{7.812e-03}, $\beta =$ \texttt{-1.974}, \texttt{9.761e-01}. Fifth cell: $\alpha =$ \texttt{1.562e-02}, \texttt{-2.980e-02}, \texttt{1.562e-02}, $\beta =$ \texttt{-1.985}, \texttt{9.886e-01}. Sixth cell: $\alpha =$ \texttt{1.866e+02}, \texttt{-3.501e+02}, \texttt{1.866e+02}, $\beta =$ \texttt{-1.979}, \texttt{9.818e-01}.}

\section{FIR filter coefficients}
\label{app:fir-coefficients}

We give here the $2N+1$ coefficients $\alpha_k$ of the causal and noncausal FIR filters used in Secs.~\ref{sec:analytic-model} and~\ref{sec:simulation}. The causal filter's $z$ transform is given by
\begin{equation*}
    \fourier{f}_\text{causal}(z) = \sum_{k=0}^{2N}{\alpha_k z^{-k}} \qc
\end{equation*}
while that of the noncausal filter is
\begin{equation*}
    \fourier{f}_\text{acausal}(z) = \sum_{k=-N}^{N}{\alpha_{N+k} z^{-k}} = z^{N} \fourier{f}_\text{causal}(z) \qs
\end{equation*}
    
Since the filter is symmetrical, the following holds
\begin{equation*}
    \alpha_k = \alpha_{2N - k} \qq{for} 0 \leq k \leq 2N \qs
\end{equation*}

We therefore give half of the coefficients $\alpha_0 \dots \alpha_n$, the rest can be deduced by symmetry: {\scriptsize \texttt{0.0225037}, \texttt{0.0224708}, \texttt{0.0223744}, \texttt{0.0222146}, \texttt{0.0219926}, \texttt{0.0217101}, \texttt{0.0213692}, \texttt{0.0209726}, \texttt{0.0205233}, \texttt{0.0200246}, \texttt{0.0194801}, \texttt{0.0188939}, \texttt{0.0182701}, \texttt{0.017613}, \texttt{0.0169273}, \texttt{0.0162175}, \texttt{0.0154884}, \texttt{0.0147444}, \texttt{0.0139904}, \texttt{0.0132308}, \texttt{0.0124699}, \texttt{0.0117121}, \texttt{0.0109613}, \texttt{0.0102213}, \texttt{0.00949556}, \texttt{0.00878736}, \texttt{0.0080996}, \texttt{0.00743485}, \texttt{0.00679539}, \texttt{0.00618313}, \texttt{0.00559967}, \texttt{0.00504627}, \texttt{0.00452385}, \texttt{0.00403305}, \texttt{0.00357419}, \texttt{0.0031473}, \texttt{0.00275216}, \texttt{0.0023883}, \texttt{0.00205505}, \texttt{0.00175152}, \texttt{0.00147665}, \texttt{0.00122927}, \texttt{0.00100804}, \texttt{0.000811548}, \texttt{0.000638312}, \texttt{0.00048679}, \texttt{0.000355413}, \texttt{0.000242603}, \texttt{0.000146789}, \texttt{6.64234e-05}, \texttt{-3.06836e-19}, \texttt{-5.39377e-05}, \texttt{-9.67836e-05}, \texttt{-0.000129861}, \texttt{-0.000154416}, \texttt{-0.000171611}, \texttt{-0.000182522}, \texttt{-0.000188139}, \texttt{-0.000189361}, \texttt{-0.000187}, \texttt{-0.000181782}, \texttt{-0.000174349}, \texttt{-0.000165266}, \texttt{-0.00015502}, \texttt{-0.000144028}, \texttt{-0.000132644}, \texttt{-0.000121158}, \texttt{-0.00010981}, \texttt{-9.87885e-05}, \texttt{-8.82397e-05}, \texttt{-7.82713e-05}, \texttt{-6.89586e-05}, \texttt{-6.03483e-05}, \texttt{-5.24636e-05}, \texttt{-4.53084e-05}, \texttt{-3.88703e-05}, \texttt{-3.31251e-05}, \texttt{-2.80388e-05}, \texttt{-2.35708e-05}, \texttt{-1.96758e-05}, \texttt{-1.63062e-05}, \texttt{-1.34132e-05}, \texttt{-1.09483e-05}, \texttt{-8.86423e-06}, \texttt{-7.11608e-06}, \texttt{-5.66147e-06}, \texttt{-4.46113e-06}, \texttt{-3.47914e-06}, \texttt{-2.68298e-06}, \texttt{-2.04359e-06}, \texttt{-1.53525e-06}, \texttt{-1.13543e-06}, \texttt{-8.24623e-07}, \texttt{-5.86081e-07}, \texttt{-4.05589e-07}, \texttt{-2.71198e-07}, \texttt{-1.72976e-07}, \texttt{-1.02753e-07}, \texttt{-5.38886e-08}, \texttt{-2.10473e-08}, \texttt{1.67089e-22}, \texttt{1.25602e-08}, \texttt{1.91793e-08}, \texttt{2.17835e-08}, \texttt{2.18024e-08}, \texttt{2.02719e-08}, \texttt{1.79219e-08}, \texttt{1.52489e-08}, \texttt{1.25738e-08}, \texttt{1.00899e-08}, \texttt{7.89963e-09}, \texttt{6.04316e-09}, \texttt{4.52036e-09}, \texttt{3.30667e-09}, \texttt{2.36459e-09}, \texttt{1.65159e-09}, \texttt{1.12523e-09}, \texttt{7.46354e-10}, \texttt{4.80713e-10}, \texttt{2.99603e-10}, \texttt{1.79833e-10}, \texttt{1.03279e-10}, \texttt{5.62201e-11}, \texttt{2.85993e-11}, \texttt{1.32841e-11}, \texttt{5.3973e-12}, \texttt{1.73705e-12}.}

\begin{acknowledgments}
This research has been supported by the Centre National d'Études Spatiales, Centrale National pour la Recherche Scientifique (CNRS) and Université Paris-Diderot. The development of \texttt{LISACode} and \texttt{LISANode} is part of the LISA Simulation Group activities. We thank Víctor Martín Hernández (IEEC, UAB) for fruitful discussions on the structure of \texttt{LISANode}. We gracefully acknowledge the suggestion by Robert Spero, Kirk McKenzie, Brent Ware, Samuel Francis and Michele Vallisneri (JPL, Caltech) that a noncausal symmetric filter would cancel the residual noise at the leading order. We are also grateful to the members of the LISA Simulation Group for their help in improving the manuscript.
\end{acknowledgments}

\bibliographystyle{apsrev4-1}
\bibliography{references}

\begin{thebibliography}{23}%
\makeatletter
\providecommand \@ifxundefined [1]{%
 \@ifx{#1\undefined}
}%
\providecommand \@ifnum [1]{%
 \ifnum #1\expandafter \@firstoftwo
 \else \expandafter \@secondoftwo
 \fi
}%
\providecommand \@ifx [1]{%
 \ifx #1\expandafter \@firstoftwo
 \else \expandafter \@secondoftwo
 \fi
}%
\providecommand \natexlab [1]{#1}%
\providecommand \enquote  [1]{``#1''}%
\providecommand \bibnamefont  [1]{#1}%
\providecommand \bibfnamefont [1]{#1}%
\providecommand \citenamefont [1]{#1}%
\providecommand \href@noop [0]{\@secondoftwo}%
\providecommand \href [0]{\begingroup \@sanitize@url \@href}%
\providecommand \@href[1]{\@@startlink{#1}\@@href}%
\providecommand \@@href[1]{\endgroup#1\@@endlink}%
\providecommand \@sanitize@url [0]{\catcode `\\12\catcode `\$12\catcode
  `\&12\catcode `\#12\catcode `\^12\catcode `\_12\catcode `\%12\relax}%
\providecommand \@@startlink[1]{}%
\providecommand \@@endlink[0]{}%
\providecommand \url  [0]{\begingroup\@sanitize@url \@url }%
\providecommand \@url [1]{\endgroup\@href {#1}{\urlprefix }}%
\providecommand \urlprefix  [0]{URL }%
\providecommand \Eprint [0]{\href }%
\providecommand \doibase [0]{http://dx.doi.org/}%
\providecommand \selectlanguage [0]{\@gobble}%
\providecommand \bibinfo  [0]{\@secondoftwo}%
\providecommand \bibfield  [0]{\@secondoftwo}%
\providecommand \translation [1]{[#1]}%
\providecommand \BibitemOpen [0]{}%
\providecommand \bibitemStop [0]{}%
\providecommand \bibitemNoStop [0]{.\EOS\space}%
\providecommand \EOS [0]{\spacefactor3000\relax}%
\providecommand \BibitemShut  [1]{\csname bibitem#1\endcsname}%
\let\auto@bib@innerbib\@empty
\bibitem [{\citenamefont {Amaro-Seoane}\ \emph {et~al.}(2017)\citenamefont
  {Amaro-Seoane} \emph {et~al.}}]{Audley:2017drz}%
  \BibitemOpen
  \bibfield  {author} {\bibinfo {author} {\bibfnamefont {P.}~\bibnamefont
  {Amaro-Seoane}} \emph {et~al.} (\bibinfo {collaboration} {LISA
  Collaboration}),\ }\href@noop {} {\emph {\bibinfo {title} {{Laser
  Interferometer Space Antenna}}}},\ \bibinfo {type} {Tech. Rep.}\ (\bibinfo
  {institution} {LISA Consortium},\ \bibinfo {year} {2017})\ \Eprint
  {http://arxiv.org/abs/1702.00786} {arXiv.org:1702.00786 [astro-ph]}
  \BibitemShut {NoStop}%
\bibitem [{\citenamefont {Armano}\ and\ \citenamefont
  {{others}}(2016)}]{Armano:2016bkm}%
  \BibitemOpen
  \bibfield  {author} {\bibinfo {author} {\bibfnamefont {M.}~\bibnamefont
  {Armano}}\ and\ \bibinfo {author} {\bibnamefont {{others}}},\ }\href@noop {}
  {\bibfield  {journal} {\bibinfo  {journal} {Phys. Rev. Lett.}\ }\textbf
  {\bibinfo {volume} {116}},\ \bibinfo {pages} {231101} (\bibinfo {year}
  {2016})}\BibitemShut {NoStop}%
\bibitem [{\citenamefont {Armano}\ \emph {et~al.}(2018)\citenamefont {Armano}
  \emph {et~al.}}]{Armano:2018hg}%
  \BibitemOpen
  \bibfield  {author} {\bibinfo {author} {\bibfnamefont {M.}~\bibnamefont
  {Armano}} \emph {et~al.},\ }\href@noop {} {\bibfield  {journal} {\bibinfo
  {journal} {Phys. Rev. Lett.}\ }\textbf {\bibinfo {volume} {120}},\ \bibinfo
  {pages} {061101} (\bibinfo {year} {2018})}\BibitemShut {NoStop}%
\bibitem [{\citenamefont {Armstrong}\ \emph {et~al.}(1999)\citenamefont
  {Armstrong}, \citenamefont {Estabrook},\ and\ \citenamefont
  {Tinto}}]{Armstrong:1999hp}%
  \BibitemOpen
  \bibfield  {author} {\bibinfo {author} {\bibfnamefont {J.~W.}\ \bibnamefont
  {Armstrong}}, \bibinfo {author} {\bibfnamefont {F.~B.}\ \bibnamefont
  {Estabrook}}, \ and\ \bibinfo {author} {\bibfnamefont {M.}~\bibnamefont
  {Tinto}},\ }\href@noop {} {\bibfield  {journal} {\bibinfo  {journal} {ApJ}\
  }\textbf {\bibinfo {volume} {527}},\ \bibinfo {pages} {814} (\bibinfo {year}
  {1999})}\BibitemShut {NoStop}%
\bibitem [{\citenamefont {Tinto}\ and\ \citenamefont
  {Armstrong}(1999)}]{Tinto:1999yr}%
  \BibitemOpen
  \bibfield  {author} {\bibinfo {author} {\bibfnamefont {M.}~\bibnamefont
  {Tinto}}\ and\ \bibinfo {author} {\bibfnamefont {J.~W.}\ \bibnamefont
  {Armstrong}},\ }\href@noop {} {\bibfield  {journal} {\bibinfo  {journal}
  {Phys. Rev.}\ }\textbf {\bibinfo {volume} {D59}},\ \bibinfo {pages} {102003}
  (\bibinfo {year} {1999})}\BibitemShut {NoStop}%
\bibitem [{\citenamefont {de~Vine}\ \emph {et~al.}(2010)\citenamefont
  {de~Vine}, \citenamefont {Ware}, \citenamefont {McKenzie}, \citenamefont
  {Spero}, \citenamefont {Klipstein},\ and\ \citenamefont
  {Shaddock}}]{deVine:2010et}%
  \BibitemOpen
  \bibfield  {author} {\bibinfo {author} {\bibfnamefont {G.}~\bibnamefont
  {de~Vine}}, \bibinfo {author} {\bibfnamefont {B.}~\bibnamefont {Ware}},
  \bibinfo {author} {\bibfnamefont {K.}~\bibnamefont {McKenzie}}, \bibinfo
  {author} {\bibfnamefont {R.~E.}\ \bibnamefont {Spero}}, \bibinfo {author}
  {\bibfnamefont {W.~M.}\ \bibnamefont {Klipstein}}, \ and\ \bibinfo {author}
  {\bibfnamefont {D.~A.}\ \bibnamefont {Shaddock}},\ }\href {\doibase
  10.1103/PhysRevLett.104.211103} {\bibfield  {journal} {\bibinfo  {journal}
  {Phys. Rev. Lett.}\ }\textbf {\bibinfo {volume} {104}},\ \bibinfo {pages}
  {211103} (\bibinfo {year} {2010})},\ \Eprint {http://arxiv.org/abs/1005.2176}
  {arXiv:1005.2176 [astro-ph.IM]} \BibitemShut {NoStop}%
\bibitem [{\citenamefont {Schwarze}\ \emph {et~al.}(2019)\citenamefont
  {Schwarze}, \citenamefont {Fernández~Barranco}, \citenamefont {Penkert},
  \citenamefont {Kaufer}, \citenamefont {Gerberding},\ and\ \citenamefont
  {Heinzel}}]{Schwarze:2018lvl}%
  \BibitemOpen
  \bibfield  {author} {\bibinfo {author} {\bibfnamefont {T.~S.}\ \bibnamefont
  {Schwarze}}, \bibinfo {author} {\bibfnamefont {G.}~\bibnamefont
  {Fernández~Barranco}}, \bibinfo {author} {\bibfnamefont {D.}~\bibnamefont
  {Penkert}}, \bibinfo {author} {\bibfnamefont {M.}~\bibnamefont {Kaufer}},
  \bibinfo {author} {\bibfnamefont {O.}~\bibnamefont {Gerberding}}, \ and\
  \bibinfo {author} {\bibfnamefont {G.}~\bibnamefont {Heinzel}},\ }\href
  {\doibase 10.1103/PhysRevLett.122.081104} {\bibfield  {journal} {\bibinfo
  {journal} {Phys. Rev. Lett.}\ }\textbf {\bibinfo {volume} {122}},\ \bibinfo
  {pages} {081104} (\bibinfo {year} {2019})},\ \Eprint
  {http://arxiv.org/abs/1810.00728} {arXiv:1810.00728 [astro-ph.IM]}
  \BibitemShut {NoStop}%
\bibitem [{\citenamefont {Otto}(2015)}]{Otto:2015wp}%
  \BibitemOpen
  \bibfield  {author} {\bibinfo {author} {\bibfnamefont {M.}~\bibnamefont
  {Otto}},\ }\emph {\bibinfo {title} {{Time-Delay Interferometry Simulations
  for the Laser Interferometer Space Antenna}}},\ \href@noop {} {Ph.D.
  thesis},\ \bibinfo  {school} {Gottfried Wilhelm Leibniz Universit{\"a}t
  Hannover} (\bibinfo {year} {2015})\BibitemShut {NoStop}%
\bibitem [{\citenamefont {Laporte}\ \emph {et~al.}(2017)\citenamefont
  {Laporte}, \citenamefont {Halloin}, \citenamefont {Br{\'e}elle},
  \citenamefont {Buy}, \citenamefont {Gr{\"u}ning},\ and\ \citenamefont
  {Prat}}]{Laporte:2017bv}%
  \BibitemOpen
  \bibfield  {author} {\bibinfo {author} {\bibfnamefont {M.}~\bibnamefont
  {Laporte}}, \bibinfo {author} {\bibfnamefont {H.}~\bibnamefont {Halloin}},
  \bibinfo {author} {\bibfnamefont {E.}~\bibnamefont {Br{\'e}elle}}, \bibinfo
  {author} {\bibfnamefont {C.}~\bibnamefont {Buy}}, \bibinfo {author}
  {\bibfnamefont {P.}~\bibnamefont {Gr{\"u}ning}}, \ and\ \bibinfo {author}
  {\bibfnamefont {P.}~\bibnamefont {Prat}},\ }\href@noop {} {\bibfield
  {journal} {\bibinfo  {journal} {J. Phys. Conf. Ser.}\ }\textbf {\bibinfo
  {volume} {840}},\ \bibinfo {pages} {012014} (\bibinfo {year}
  {2017})}\BibitemShut {NoStop}%
\bibitem [{\citenamefont {Gr{\"u}ning}\ \emph {et~al.}(2015)\citenamefont
  {Gr{\"u}ning} \emph {et~al.}}]{Gruning:2015cp}%
  \BibitemOpen
  \bibfield  {author} {\bibinfo {author} {\bibfnamefont {P.}~\bibnamefont
  {Gr{\"u}ning}} \emph {et~al.},\ }\href@noop {} {\bibfield  {journal}
  {\bibinfo  {journal} {Exp. Astron.}\ }\textbf {\bibinfo {volume} {39}},\
  \bibinfo {pages} {281} (\bibinfo {year} {2015})}\BibitemShut {NoStop}%
\bibitem [{\citenamefont {Cruz}\ \emph {et~al.}(2006)\citenamefont {Cruz},
  \citenamefont {Thorpe}, \citenamefont {Hartman},\ and\ \citenamefont
  {Mueller}}]{Cruz:2006js}%
  \BibitemOpen
  \bibfield  {author} {\bibinfo {author} {\bibfnamefont {R.~J.}\ \bibnamefont
  {Cruz}}, \bibinfo {author} {\bibfnamefont {J.~I.}\ \bibnamefont {Thorpe}},
  \bibinfo {author} {\bibfnamefont {M.}~\bibnamefont {Hartman}}, \ and\
  \bibinfo {author} {\bibfnamefont {G.}~\bibnamefont {Mueller}},\ }\bibfield
  {booktitle} {\emph {\bibinfo {booktitle} {{Proceedings, 6th International
  LISA Symposium on Laser interferometer space antenna: Greenbelt, USA, June
  19-23, 2006}}},\ }\href {\doibase 10.1063/1.2405062} {\bibfield  {journal}
  {\bibinfo  {journal} {AIP Conf. Proc.}\ }\textbf {\bibinfo {volume} {873}},\
  \bibinfo {pages} {319} (\bibinfo {year} {2006})}\BibitemShut {NoStop}%
\bibitem [{\citenamefont {Vallisneri}(2005)}]{Vallisneri:2005ca}%
  \BibitemOpen
  \bibfield  {author} {\bibinfo {author} {\bibfnamefont {M.}~\bibnamefont
  {Vallisneri}},\ }\href {\doibase 10.1103/PhysRevD.71.022001} {\bibfield
  {journal} {\bibinfo  {journal} {Phys. Rev. D}\ }\textbf {\bibinfo {volume}
  {71}},\ \bibinfo {pages} {022001} (\bibinfo {year} {2005})},\ \Eprint
  {http://arxiv.org/abs/gr-qc/0407102} {arXiv:gr-qc/0407102} \BibitemShut
  {NoStop}%
\bibitem [{\citenamefont {Petiteau}(2008)}]{Petiteau:2008wa}%
  \BibitemOpen
  \bibfield  {author} {\bibinfo {author} {\bibfnamefont {A.}~\bibnamefont
  {Petiteau}},\ }\emph {\bibinfo {title} {{De la simulation de LISA {\`a}
  l'analyse des donn{\'e}es}}},\ \href@noop {} {Ph.D. thesis},\ \bibinfo
  {school} {Universit{\'e} Paris-Diderot} (\bibinfo {year} {2008})\BibitemShut
  {NoStop}%
\bibitem [{\citenamefont {Petiteau}\ \emph {et~al.}(2008)\citenamefont
  {Petiteau}, \citenamefont {Auger}, \citenamefont {Halloin}, \citenamefont
  {Jeannin}, \citenamefont {Plagnol}, \citenamefont {Pireaux}, \citenamefont
  {Regimbau},\ and\ \citenamefont {Vinet}}]{Petiteau:2008ke}%
  \BibitemOpen
  \bibfield  {author} {\bibinfo {author} {\bibfnamefont {A.}~\bibnamefont
  {Petiteau}}, \bibinfo {author} {\bibfnamefont {G.}~\bibnamefont {Auger}},
  \bibinfo {author} {\bibfnamefont {H.}~\bibnamefont {Halloin}}, \bibinfo
  {author} {\bibfnamefont {O.}~\bibnamefont {Jeannin}}, \bibinfo {author}
  {\bibfnamefont {E.}~\bibnamefont {Plagnol}}, \bibinfo {author} {\bibfnamefont
  {S.}~\bibnamefont {Pireaux}}, \bibinfo {author} {\bibfnamefont
  {T.}~\bibnamefont {Regimbau}}, \ and\ \bibinfo {author} {\bibfnamefont
  {J.-Y.}\ \bibnamefont {Vinet}},\ }\href {\doibase 10.1103/PhysRevD.77.023002}
  {\bibfield  {journal} {\bibinfo  {journal} {Phys. Rev. D}\ }\textbf {\bibinfo
  {volume} {77}},\ \bibinfo {pages} {023002} (\bibinfo {year} {2008})},\
  \Eprint {http://arxiv.org/abs/0802.2023} {arXiv:0802.2023 [gr-qc]}
  \BibitemShut {NoStop}%
\bibitem [{\citenamefont {Bayle}\ \emph {et~al.}(2019)\citenamefont {Bayle},
  \citenamefont {Lilley},\ and\ \citenamefont {Petiteau}}]{Bayle:tbp}%
  \BibitemOpen
  \bibfield  {author} {\bibinfo {author} {\bibfnamefont {J.-B.}\ \bibnamefont
  {Bayle}}, \bibinfo {author} {\bibfnamefont {M.}~\bibnamefont {Lilley}}, \
  and\ \bibinfo {author} {\bibfnamefont {A.}~\bibnamefont {Petiteau}},\
  }\href@noop {} {\enquote {\bibinfo {title} {{LISANode: A modular simulator
  for LISA}},}\ } (\bibinfo {year} {2019}),\ \bibinfo {note} {to be
  published}\BibitemShut {NoStop}%
\bibitem [{\citenamefont {Otto}\ \emph {et~al.}(2012)\citenamefont {Otto},
  \citenamefont {Heinzel},\ and\ \citenamefont {Danzmann}}]{Otto:2012dk}%
  \BibitemOpen
  \bibfield  {author} {\bibinfo {author} {\bibfnamefont {M.}~\bibnamefont
  {Otto}}, \bibinfo {author} {\bibfnamefont {G.}~\bibnamefont {Heinzel}}, \
  and\ \bibinfo {author} {\bibfnamefont {K.}~\bibnamefont {Danzmann}},\
  }\href@noop {} {\bibfield  {journal} {\bibinfo  {journal} {Class. Quant.
  Grav.}\ }\textbf {\bibinfo {volume} {29}},\ \bibinfo {pages} {205003}
  (\bibinfo {year} {2012})}\BibitemShut {NoStop}%
\bibitem [{\citenamefont {Tinto}\ and\ \citenamefont
  {Hartwig}(2018)}]{Tinto:2018kij}%
  \BibitemOpen
  \bibfield  {author} {\bibinfo {author} {\bibfnamefont {M.}~\bibnamefont
  {Tinto}}\ and\ \bibinfo {author} {\bibfnamefont {O.}~\bibnamefont
  {Hartwig}},\ }\href {\doibase 10.1103/PhysRevD.98.042003} {\bibfield
  {journal} {\bibinfo  {journal} {Phys. Rev. D}\ }\textbf {\bibinfo {volume}
  {98}},\ \bibinfo {pages} {042003} (\bibinfo {year} {2018})},\ \Eprint
  {http://arxiv.org/abs/1807.02594} {arXiv:1807.02594 [gr-qc]} \BibitemShut
  {NoStop}%
\bibitem [{\citenamefont {Barke}(2015)}]{Barke:2015wb}%
  \BibitemOpen
  \bibfield  {author} {\bibinfo {author} {\bibfnamefont {S.}~\bibnamefont
  {Barke}},\ }\emph {\bibinfo {title} {{Inter-Spacecraft Frequency Distribution
  for Future Gravitational Wave Observatories}}},\ \href@noop {} {Ph.D.
  thesis},\ \bibinfo  {school} {Gottfried Wilhelm Leibniz Universit{\"a}t
  Hannover} (\bibinfo {year} {2015})\BibitemShut {NoStop}%
\bibitem [{\citenamefont {Dhurandhar}\ \emph {et~al.}(2002)\citenamefont
  {Dhurandhar}, \citenamefont {Rajesh~Nayak},\ and\ \citenamefont
  {Vinet}}]{Dhurandhar:2002zcl}%
  \BibitemOpen
  \bibfield  {author} {\bibinfo {author} {\bibfnamefont {S.}~\bibnamefont
  {Dhurandhar}}, \bibinfo {author} {\bibfnamefont {K.}~\bibnamefont
  {Rajesh~Nayak}}, \ and\ \bibinfo {author} {\bibfnamefont {J.}~\bibnamefont
  {Vinet}},\ }\href {\doibase 10.1103/PhysRevD.65.102002} {\bibfield  {journal}
  {\bibinfo  {journal} {Phys. Rev. D}\ }\textbf {\bibinfo {volume} {65}},\
  \bibinfo {pages} {102002} (\bibinfo {year} {2002})},\ \Eprint
  {http://arxiv.org/abs/gr-qc/0112059} {arXiv:gr-qc/0112059} \BibitemShut
  {NoStop}%
\bibitem [{\citenamefont {Rajesh~Nayak}\ \emph {et~al.}(2006)\citenamefont
  {Rajesh~Nayak}, \citenamefont {Koshti}, \citenamefont {Dhurandhar},\ and\
  \citenamefont {Vinet}}]{Nayak:2006zm}%
  \BibitemOpen
  \bibfield  {author} {\bibinfo {author} {\bibfnamefont {K.}~\bibnamefont
  {Rajesh~Nayak}}, \bibinfo {author} {\bibfnamefont {S.}~\bibnamefont
  {Koshti}}, \bibinfo {author} {\bibfnamefont {S.~V.}\ \bibnamefont
  {Dhurandhar}}, \ and\ \bibinfo {author} {\bibfnamefont {J.-Y.}\ \bibnamefont
  {Vinet}},\ }\href@noop {} {\bibfield  {journal} {\bibinfo  {journal} {Class.
  Quant. Grav.}\ }\textbf {\bibinfo {volume} {23}},\ \bibinfo {pages} {1763}
  (\bibinfo {year} {2006})}\BibitemShut {NoStop}%
\bibitem [{\citenamefont {Dhurandhar}\ \emph {et~al.}(2004)\citenamefont
  {Dhurandhar}, \citenamefont {Nayak}, \citenamefont {Koshti},\ and\
  \citenamefont {Vinet}}]{Dhurandhar:2004jr}%
  \BibitemOpen
  \bibfield  {author} {\bibinfo {author} {\bibfnamefont {S.~V.}\ \bibnamefont
  {Dhurandhar}}, \bibinfo {author} {\bibfnamefont {K.~R.}\ \bibnamefont
  {Nayak}}, \bibinfo {author} {\bibfnamefont {S.}~\bibnamefont {Koshti}}, \
  and\ \bibinfo {author} {\bibfnamefont {J.-Y.}\ \bibnamefont {Vinet}},\
  }\href@noop {} {\bibfield  {journal} {\bibinfo  {journal} {Class. Quant.
  Grav.}\ ,\ \bibinfo {pages} {481}} (\bibinfo {year} {2004})},\ \Eprint
  {http://arxiv.org/abs/gr-qc/0410093} {arXiv:gr-qc/0410093 [gr-qc]}
  \BibitemShut {NoStop}%
\bibitem [{\citenamefont {{LISA Science Study Team}}(2018)}]{MRDv1}%
  \BibitemOpen
  \bibfield  {author} {\bibinfo {author} {\bibnamefont {{LISA Science Study
  Team}}},\ }\href@noop {} {\emph {\bibinfo {title} {{LISA Science Requirements
  Document}}}},\ \bibinfo {type} {Tech. Rep.}\ \bibinfo {number} {1.0}\
  (\bibinfo  {institution} {ESA},\ \bibinfo {year} {2018})\ \bibinfo {note}
  {\url{https://www.cosmos.esa.int/web/lisa/lisa-documents/}}\BibitemShut
  {NoStop}%
\bibitem [{\citenamefont {Babak}\ \emph {et~al.}(2008)\citenamefont {Babak},
  \citenamefont {Baker} \emph {et~al.}}]{Babak:2008kx}%
  \BibitemOpen
  \bibfield  {author} {\bibinfo {author} {\bibfnamefont {S.}~\bibnamefont
  {Babak}}, \bibinfo {author} {\bibfnamefont {J.~G.}\ \bibnamefont {Baker}},
  \emph {et~al.},\ }\href@noop {} {\bibfield  {journal} {\bibinfo  {journal}
  {Class. Quantum Grav.}\ }\textbf {\bibinfo {volume} {25}},\ \bibinfo {pages}
  {184026} (\bibinfo {year} {2008})}\BibitemShut {NoStop}%
\end{thebibliography}%

\end{document}